\documentclass[journal]{IEEEtran}

\usepackage{amsmath,amssymb,amsfonts}
\usepackage{algorithmic}
\usepackage{graphicx}
\usepackage{textcomp}
\usepackage{array}
\usepackage{booktabs}
\usepackage{multirow}
\usepackage{url}
\usepackage[T1]{fontenc}

\newcommand{\um}{\,$\mu$m}

\newcolumntype{C}[1]{>{\centering\arraybackslash}p{#1}}
\newtheorem{proposition}{Proposition}
\newtheorem{remark}{Remark}

\graphicspath{{simulationResults/}}

\begin{document}

\title{Toward Robotic Neural-Thread Insertion: Feasibility-Preserving Relative-Motion Control on a Pulsating Cortical Surface}

\author{Yongyan~Cao and Xiaobo Li
\thanks{Y. Cao is with Voryx Robotics LLC, San Jose, CA 95110 USA
(e-mail: yongyancao@gmail.com).}}

\maketitle

\begin{abstract}
Flexible neural electrode threads must be placed at a prescribed depth while the cortical surface moves with cardiac and respiratory pulsation. A controller tracking a fixed point in the laboratory frame cannot distinguish commanded insertion from tissue motion; the error appears as both a depth offset and relative tip--tissue velocity during contact. This paper formulates thread insertion in tissue-relative coordinates: a harmonic observer predicts delayed cortical-surface motion over the control horizon, a constrained model-predictive controller regulates the tip about that prediction under actuator and coupled lateral-velocity limits, and an augmented disturbance state removes the steady offset from persistent contact force and model mismatch.

The primary result is that the coupled lateral-velocity (shear) constraint must be made feasibility-preserving to be usable: at 10\um\ RMS per-axis sensing noise, a shared-slack relaxation holds the 0.80~mm/s shear budget on 10 of 10 seeds, while the matched hard constraint loses one seed to a persistently infeasible horizon and an approximately 308~mm/s excursion and a cost-only controller violates on 10 of 10. That ranking is unchanged when the benchmark is re-run at directly measured small-animal cortical amplitudes roughly one sixth of the stress envelope, because at this sensing noise level the shear is noise-driven rather than motion-driven.

Regulation is offset-free: contact-window RMS relative-placement error is 1.1\um, against 76.0\um\ for delayed-feedback impedance, at the cost of higher peak contact force (1.47 vs.\ 1.14~mN) since the controller drives to commanded depth rather than yielding. A two-vertex Lyapunov certificate for the finite-horizon gain holds over $-40\%/{+}50\%$ reflected-mass mismatch and, measured over the benchmark, the constraints it assumes inactive are inactive on 94\% of control ticks and throughout both measurement windows. This is a simulation-based control benchmark with a rigid tip; Section~\ref{sec:scope} states the scope in full.
\end{abstract}


\section{Introduction}
\IEEEPARstart{F}{lexible} neural electrode threads must be placed at a prescribed depth while the tissue they enter is not stationary. During an open-cranium procedure, cardiac and respiratory motion displace the cortical surface throughout the interval in which the robot approaches and penetrates it. The state that matters is therefore not the tool's position in the room, but its position and velocity \emph{relative to the local tissue surface}.

Let $p_n$ denote the thread-tip position along the insertion axis and $d(t)$ the displacement of the tissue surface at the target site. The relevant coordinate is the gap
\begin{equation}
r = p_n - d.
\end{equation}
Writing the problem this way exposes two errors that a fixed lab-frame target hides. First, if the surface rises by $\Delta$ at the instant of placement, a controller that continues to track a fixed room-frame point realizes an insertion depth shifted by the same amount; reported cortical-surface motion ranges from tens of microns in anesthetized small animals \cite{gilletti2006,nomura2024} to several hundred microns after human craniotomy \cite{faria2014}, in either case comparable to or larger than the placement tolerance of neural-thread insertion (Section~\ref{sec:amplitudes}). Second, any mismatch between $\dot p_n$ and $\dot d$ produces relative motion during contact. We treat this relative velocity, and especially its lateral component $\dot{\mathbf p}_{n,\mathrm{lat}}-\dot{\mathbf d}_{\mathrm{lat}}$, as a mechanical surrogate for tip--tissue shear -- a quantity the controller can measure and bound. Cardiac and respiratory motion is three-dimensional, producing lateral slip as well as normal excursion, so an axial-only controller cannot reduce this second error even if it eliminates the first.

Physiological motion is not perfectly periodic, but most of its energy sits near the cardiac and respiratory frequencies. That structure makes short-horizon prediction useful: instead of reacting only to a delayed surface measurement, the controller can estimate the current phase of the motion, propagate the estimate through the sensing delay, and predict the surface trajectory over the control horizon. The insertion target then moves with the tissue rather than staying fixed in the room.

The resulting control problem has three parts: track a commanded depth relative to the moving surface, reject the persistent load that appears after contact, and respect limits on actuation and relative velocity. We address these with a harmonic motion observer, an augmented disturbance estimate, and a constrained MPC formulated directly in tissue-relative coordinates.

\subsection{Contributions}
Preview-based motion compensation and offset-free MPC are established tools, combined for other organs (Section~\ref{sec:relwork}). What is new here is not accuracy on a motion-compensation task but the behavior of the \emph{safety} constraint that this task requires: a bound on lateral tip--tissue velocity couples the control axes, and the natural hard-constrained formulation fails exactly when it is needed. The contributions, in the order we would defend them:
\begin{enumerate}
\item \textbf{A feasibility-preserving coupled shear constraint.} The three-axis controller bounds lateral relative velocity as a coupled norm, $\|\dot{\mathbf r}_{\mathrm{lat}}\|\le v_{\mathrm{shear}}$, rather than per axis. Imposed as hard polyhedral rows, that bound conflicts with the horizon state under degraded sensing and the QP becomes infeasible; a shared nonnegative slack -- one per prediction step, common to all facets, so the relaxation acts on the coupled geometry and not on a single Cartesian axis -- restores solvability by construction. Measured at 10\um\ RMS per-axis sensing noise, this is the difference between 0 of 10 seeds violating the budget and a hard-constrained run that loses a seed to a persistently infeasible horizon and an approximately 308~mm/s excursion, while a cost-only controller violates on 10 of 10 (Sections~\ref{sec:multidof}, \ref{app:multidof}).
\item \textbf{A tissue-relative formulation} of flexible-thread insertion that treats depth error and lateral relative velocity within the same control model, rather than as separate tracking and safety problems, so that one controller covers free-space approach and post-contact hold.
\item \textbf{Offset-free regulation about a previewed moving equilibrium.} A harmonic observer estimates latency-delayed cardiac and respiratory motion, and an integrating acceleration-bias state removes the depth offset that the persistent contact load otherwise leaves -- 76.0 to 1.1\um\ contact-window RMS against the delayed-feedback baseline, without a force sensor.
\item \textbf{An evaluation reported across two motion-amplitude regimes.} The benchmark is run both at a conservative stress envelope near the upper bound of published human open-craniotomy cortical motion and at directly measured small-animal cortical amplitudes roughly one sixth of it (Section~\ref{sec:amplitudes}), so a reader can see which conclusions are amplitude-dependent and which are not, together with a corrected 10\um\ sensor-noise study, parameter mismatch, broadband disturbance, cardiac-rate drift, a $2\times2$ mechanism ablation over $N=30$ seeds, and a measurement of how often the stability result's constraint-inactive hypothesis actually holds.
\end{enumerate}

The control ingredients are introduced after the biomedical task and metrics are defined.

\subsection{Scope and Limitations}
\label{sec:scope}
The scope of the study is stated here once, and not repeated after each result. This is a simulation-based control benchmark in MuJoCo. The modeled tip is a rigid contact point on the insertion axis: the flexible thread, its carrier needle, and the release mechanics of an actual inserter are not modeled, so the paper is about relative-motion regulation for an insertion tool, not about thread mechanics. Lateral relative velocity is used as a \emph{mechanical} surrogate for tip--tissue shear (Section~\ref{sec:surrogate}); its connection to vascular damage, inflammation, or chronic thread drift is an experimental question this paper does not answer, and the claim made is only that the simulations reduce relative placement error and relative velocity in a pulsating-tissue model. The two motion-amplitude regimes are bounded by the published measurements collected in Section~\ref{sec:amplitudes}, none of which is a species- and procedure-matched cortical measurement for a thread-insertion robot. The formal stability result (Proposition~\ref{prop:iss}) is proved for the constraint-inactive feedback law, not for the constrained receding-horizon QP the benchmarks run; Remark~\ref{rem:constrained-gap} states the gap and Section~\ref{sec:activity} measures how much of the benchmark falls inside the proved case. The contact-force constraint caps a model-predicted force, not a validated measurement (Appendix~\ref{sec:force}), and solver timing is measured for the QP alone, not for a deployed control cycle on embedded hardware. Appendix~\ref{sec:limitations} lists the future work that would close each of these.

\section{Related Work and Comparison Scope}
\label{sec:relwork}
Physiological-motion compensation has a long history in robotic surgery: predictive filtering and heartbeat synchronization were introduced to stabilize tools relative to moving organs \cite{ginhoux2005,nakamura2001}, and later work combined motion prediction with force feedback, viscoelastic contact models, predictive control under delayed imaging, and switched impedance control \cite{cagneau2007,moreira2014,bowthorpe2016,cheng2018} -- \cite{moreira2014} reports 87\%/79\% breathing/cardiac compensation ratios. These studies show that prediction and compliant contact regulation combine well.

Two studies target intracranial motion directly, compensating micron-scale ($1.7$--$3.7$\um) patch-pipette motion during in-vivo recording: Stoy et al.\ using ECG and respiratory signals \cite{stoy2020}, and Zhang et al.\ estimating physiological motion from electrical bio-impedance and compensating it with an extended Kalman filter \cite{zhang2023}. Both hold a pipette steady for recording rather than controlling insertion depth, and neither formulates depth relative to a moving cortical surface or imposes a lateral relative-velocity constraint during penetration.

Robotic needle and thread manipulation addresses related but distinct tasks. A bilinear MPC steers a tendon-driven brachytherapy stylet toward planned, not physiological, moving targets at millimeter accuracy ($\sim$1.45~mm fixed, 8.3~mm moving) \cite{kheradmand2026}, without impedance rendering or a relative-velocity constraint. Iterative learning control improves rotational needle insertion for subretinal injection over repeated trials \cite{foroutani2025} -- complementary to, rather than competing with, a single-shot law like this one. Closest in spirit is deep-learning-based retinal motion compensation for subretinal injection \cite{wu2025retinal}: an LSTM predicts membrane displacement from intraoperative OCT and synchronizes needle motion to sub-16.4\um\ tracking error, but it targets retinal rather than cortical placement and reports neither a 3-D coupled shear constraint nor offset-free contact regulation. Vision-language-action suturing \cite{haworth2025} optimizes insertion-point targeting rather than model-based contact regulation; tactile-guided RL needle-threading \cite{yu2023} targets fine manipulation against a rigid, stationary eyelet rather than moving tissue; and velocity modulation to bound tissue trauma also appears in electrosurgery \cite{riaziat2025}, without insertion or motion compensation. None of these reports relative-depth regulation about a 3-D pulsating equilibrium with a lateral-shear metric, which is the benchmark established here.

The contribution is a task formulation and controller for flexible neural-thread insertion in which the reference frame moves with the tissue and lateral relative velocity is constrained explicitly. Our own prior work \cite{cao2026interaction} established the constant-$A_d$/affine-$B_d(\rho)$ backbone and its two-vertex robustness certificate for a fixed task-space reference; this paper specializes that backbone to a moving, physiological-preview-driven equilibrium (Section~\ref{sec:offline}). Impedance control \cite{hogan1985}, output regulation and adaptive sinusoidal rejection \cite{francis1976,bodson1997}, and offset-free MPC / extended-state-observer ideas \cite{pannocchia2003,han2009} supply the remaining classical foundations. The physiological-motion predictor of Section~\ref{sec:physio} is itself a model-based alternative to learned periodic-motion representations such as movement primitives \cite{gutierrez2026}: the exosystem trades the flexibility of learning from demonstration for an explicit preview signal and the ISS guarantee of Proposition~\ref{prop:iss}, at the cost of assuming a harmonic disturbance form.

We found no prior work reporting this exact benchmark -- relative-depth insertion into a 3-D pulsating cortical surface under an explicit lateral-shear constraint -- so Section~\ref{sec:sim} compares the proposed controller against internal baselines (lab-frame tracking, delayed-feedback impedance, axial-only and fixed-frequency preview) and, more informatively, against ablations of itself, rather than against published insertion hardware. Table~\ref{tab:compare} places the closest external methods by capability; their reported accuracies come from different tasks and experimental settings and are not commensurable with ours, which is also why we do not report a reimplementation of one of them: a method retuned by us, on a plant and task it was not designed for, would produce a number that says more about our tuning than about the method.

\begin{table*}[!t]
\renewcommand{\arraystretch}{1.3}
\caption{Qualitative Capability Comparison with Adjacent Robotic-Insertion Control Methods. Accuracies Are on Each Method's Own Task/Benchmark and Are Not Directly Commensurable}
\label{tab:compare}
\centering
\footnotesize
\begin{tabular}{p{2.5cm}p{2.6cm}C{1.25cm}C{1.4cm}C{1.35cm}C{1.5cm}p{2.3cm}}
\toprule
Method & Task / plant & Physio.\ motion comp. & Impedance / contact reg. & Shear (rel.-vel.) constr. & Real-time vs iterative & Reported accuracy \\
\midrule
\textbf{Preview relative-motion control (this work)} & relative-depth insertion-tool regulation, 3-D pulsating cortex (MuJoCo) & \checkmark\ periodic-motion preview & \checkmark\ bias compensation & \checkmark\ coupled lateral & \checkmark\ 1~kHz QP & 1.0--12\,\textmu m relative placement; feasible under degraded sensing (sim) \\
Bilinear MPC steerable stylet \cite{kheradmand2026} & needle steering to target, brachytherapy phantom & --- planned target & --- & --- & \checkmark\ online MPC & $\sim$1.45~mm fixed, 8.3~mm moving (phantom) \\
Iterative learning control \cite{foroutani2025} & rotational needle penetration, subretinal injection & --- & partial (force $\downarrow$) & --- & iterative (multi-trial) & success-rate, ex-vivo pig eye \\
Thermal-imaging velocity control \cite{riaziat2025} & electrosurgical tissue cutting & --- & --- & velocity--damage (thermal) & \checkmark\ online & 3$\times$ success, 2$\times$ peak force $\downarrow$ \\
Vision-language-action policy \cite{haworth2025} & autonomous suturing insertion & --- & --- & --- & learned policy & 59--74\% insertion-point targeting gain \\
Viscoelastic force control \cite{moreira2014} & soft-tissue contact, physiological motion (organ, phantom/ex-vivo) & \checkmark\ breathing/cardiac & \checkmark\ viscoelastic model & --- & \checkmark\ online & 87\%/79\% breathing/cardiac compensation ratio \\
\bottomrule
\end{tabular}
\end{table*}
The symbols used throughout are summarized below:
\begin{itemize}
\item $p_n,\dot p_n$ -- needle/electrode-tip position, velocity along the insertion axis.
\item $d(t),\dot d(t)$ -- cortical-surface displacement / rate at the target site.
\item $r=p_n-d$ -- tip position relative to tissue (gap; negative $=$ penetration).
\item $r_{\mathrm{ref}}(t)$ -- commanded relative profile; $e=r_{\mathrm{ref}}-r$ regulated error.
\item $w$ -- exosystem state, $\dot w=Sw,\ d=h^\top w$.
\item $\tau$ -- cortical-surface sensing latency.
\item $u$ -- insertion-axis command (force); $m$ effective tip mass; $\theta=1/m$.
\item $F_{\mathrm{ext}}$ -- tissue reaction (contact) force.
\item $A_d,B_d,E_d$ -- discrete state / input / disturbance matrices, period $T_s$.
\item $N$ -- prediction (preview) horizon.
\item $\hat d^{\mathrm{surf}}_{k+i}$ -- latency-compensated surface preview, with ride velocity $\dot{\hat d}^{\mathrm{surf}}_{k+i}$.
\item $\hat a$ -- augmented acceleration-bias estimate for offset-free contact / model-error rejection.
\end{itemize}

\section{System Model}
\label{sec:sysmodel}

\subsection{Insertion-Axis Dynamics}
\label{sec:axis}
After inner-loop computed-torque/servo compensation of gravity and Coriolis terms, the insertion axis reduces to a double integrator driven by the command and, during contact, perturbed by the tissue reaction force:
\begin{equation}
\dot x_n=\underbrace{\begin{bmatrix}0&1\\0&0\end{bmatrix}}_{A_c}x_n
+\underbrace{\begin{bmatrix}0\\1/m(\rho)\end{bmatrix}}_{B_c(\rho)}u
+\underbrace{\begin{bmatrix}0\\1/m(\rho)\end{bmatrix}}_{E_c(\rho)}F_{\mathrm{ext}},
\end{equation}
with $x_n=[p_n,\ \dot p_n]^\top$.

$A_c$ is the pure integrator chain: it carries no physical parameter. All parameter dependence -- the effective tip mass, which changes between free-space approach and tissue penetration -- sits in the input/disturbance column. Exact zero-order-hold discretization at $T_s$ therefore gives a \emph{constant} state matrix and a \emph{parameter-affine} input matrix:
\begin{equation}
\begin{aligned}
&A_d=\begin{bmatrix}1&T_s\\0&1\end{bmatrix},\quad
B_d(\theta)=\theta\,\bar B,\\
&\bar B=\begin{bmatrix}T_s^2/2\\ T_s\end{bmatrix},\quad \theta=\tfrac1m\in[\underline\theta,\overline\theta].
\end{aligned}
\end{equation}
This separation is useful computationally: the prediction matrices built from $A_d$ can be formed once, while changes in reflected inertia enter through a single scalar scheduling parameter. It also makes a common quadratic stability certificate over a bounded mass interval checkable at the two interval endpoints, as used in Section~\ref{sec:offline}.

\subsection{Physiological-Motion Predictor}
\label{sec:physio}
The cortical displacement is modeled as the output of a marginally stable linear exosystem whose modes are the physiological frequencies and their harmonics \cite{francis1976,bodson1997}:
\begin{equation}
\dot w = S\,w,\quad d = h^\top w,\quad
S=\mathrm{blkdiag}\!\big(0,S_{\omega_c},S_{2\omega_c},S_{\omega_r}\big),
\end{equation}
with $S_{\omega}=\left[\begin{smallmatrix}0&\omega\\ -\omega&0\end{smallmatrix}\right]$, equivalently $d(t)=a_0+\sum_i A_i\sin(\omega_i t+\phi_i)$ with slowly drifting amplitudes and phases. The state $w$ carries a constant component, the cardiac fundamental, its second harmonic, and a respiratory component.

Once $w_k$ is estimated, the entire future periodic surface motion follows from linear propagation:
\begin{equation}
\hat d_{k+i}=h^\top e^{S iT_s}\hat w_k.
\end{equation}
This exosystem is the formal carrier of preview: estimating $w_k$ gives infinite preview of the periodic component at the cost of one linear propagation.

Cardiac and respiratory frequencies can be supplied by auxiliary channels (ECG, a respiration belt) or estimated online, and because anaesthesia moves heart rate intra-operatively, they must be tracked online rather than assumed fixed. A stale $\omega_c$ under sparse sensing can make the preview actively harmful; Section~\ref{sec:freqadapt} quantifies this failure and introduces a frequency-adaptive observer that prevents it.

For three-dimensional motion, each Cartesian component $d_x,d_y,d_z$ carries its own exosystem, sharing the physiological frequencies but with component-specific amplitude and phase -- the cardiac line drives both axial expansion and lateral slip. The lateral channels are what the multi-axis controller of Section~\ref{sec:multidof} previews to suppress shear.

\subsection{Tissue Contact}
Cortical tissue (Young's modulus $\sim$1--10~kPa \cite{budday2017}) is modeled, for controller synthesis, as a Kelvin--Voigt viscoelastic port about its moving equilibrium $d(t)$:
\begin{equation}
F_{\mathrm{ext}} = -K_{\mathrm{env}}\,(p_n-d) - B_{\mathrm{env}}\,(\dot p_n-\dot d),\quad (p_n-d)<0,
\end{equation}
active only once the tip has penetrated. This model motivates the contact-disturbance channel and the force-constrained extensions of Section~\ref{sec:force}; the MuJoCo experiments generate contact directly through the simulator's own soft-contact model rather than evaluating this equation, and read the resulting force back as the safety-relevant quantity.

\subsection{Why Lateral Relative Velocity Is the Shear Surrogate}
\label{sec:surrogate}
Tip--tissue shear is not directly observable, so the controller must bound a proxy, and the choice deserves to be stated rather than assumed. At least four candidates exist: instantaneous lateral relative velocity $\|\dot{\mathbf r}_{\mathrm{lat}}\|$; lateral relative displacement $\|\mathbf r_{\mathrm{lat}}\|$; accumulated slip path length $\int\|\dot{\mathbf r}_{\mathrm{lat}}\|\,dt$; and tissue shear strain or strain rate at the contact patch.

We bound velocity, for two reasons. First, it is the only one of the four that is a linear function of the same predicted state the tracking cost already uses, so a coupled Euclidean bound becomes a set of polyhedral rows in the existing QP at no modeling cost -- and, being a rate, it stays meaningful when the reference itself is moving, which a displacement bound about a pulsating equilibrium does not. Second, velocity-limited advance has precedent as a trauma-limiting control variable in tissue-cutting work \cite{riaziat2025}.

The limitation this creates should be read plainly: \emph{peak velocity does not bound accumulated slip}. A slow, sustained drag can accumulate a large relative displacement while never approaching the velocity budget, and conversely a brief transient can exceed the budget while displacing the tissue negligibly. Tissue injury is, on the available biomechanical evidence, more closely related to strain and to its accumulation than to instantaneous rate, so a velocity budget is a controllable surrogate for a quantity we cannot yet measure, not a damage model. Two things partially cover the gap in what follows: lateral relative \emph{displacement} is reported alongside velocity for every 3-DOF result, and violation \emph{duration}, not only violation occurrence, is logged in the noise sweep (Appendix~\ref{app:multidof}). An accumulated-slip row, $\sum_i \|\dot{\mathbf r}_{\mathrm{lat},i}\|T_s \le s_{\max}$, is expressible in the same polyhedral form as the velocity rows and would be the natural next constraint once a slip or strain threshold has been measured; it is not implemented here.

\subsection{Combined Plant in Tissue-Relative Coordinates}
Augmenting the robot state with the exosystem and writing the regulated error turns the problem into standard output regulation:
\begin{align}
\begin{bmatrix}\dot x_n\\ \dot w\end{bmatrix}
&=\begin{bmatrix}A_c & 0\\ 0 & S\end{bmatrix}\begin{bmatrix}x_n\\ w\end{bmatrix}
+\begin{bmatrix}B_c\\ 0\end{bmatrix}u,\\
e &= r_{\mathrm{ref}} - \big([1\ 0]x_n - h^\top w\big).
\end{align}
This change of coordinates is the central modeling step: in laboratory coordinates surface motion is an external disturbance, while in tissue-relative coordinates it becomes part of the reference trajectory, and insertion depth carries the same meaning throughout approach and contact.

\section{Control Objective: Follow the Moving Tissue Surface}
\label{sec:impedance}
Insertion has two phases that must be handled by one controller: during free-space approach the tip should follow $d(t)$ closely, where a relatively high apparent stiffness is acceptable; during penetration it should render compliance so that $F_{\mathrm{ext}}$ stays bounded while still following slow tissue motion. Both phases are the same object once motion compensation is read as impedance control about the moving equilibrium $d(t)$. The textbook impedance target
\begin{equation}
M_m\,\ddot{\tilde r}+B_m\,\dot{\tilde r}+K_m\,\tilde r=F_{\mathrm{ext}},\qquad \tilde r = r-r_{\mathrm{ref}},
\end{equation}
renders, in the unconstrained disturbance-free limit, the same second-order port the MPC realizes; the impedance gains are the closed-loop gains induced by the Riccati solution, not the cost weights themselves (Section~\ref{sec:equiv}). Scheduling those weights shifts the rendered port from a stiff tracker in approach to a compliant follower in contact -- Section~\ref{sec:impedance-family} benchmarks that schedule rather than only asserting it, and finds it buys a 28\% reduction in peak contact force for 0.9\um\ of placement error.

The benchmarked controller sits at the rigid end of this family, however, not at a literal finite-$K_m$ compliant law. For a genuinely finite $K_m$, a persistent contact force produces a nonzero steady-state deflection $\tilde r_{\mathrm{ss}}=F_{\mathrm{ext}}/K_m$. The offset-free acceleration-bias observer of Section~\ref{sec:architecture} instead treats the persistent contact reaction as a disturbance to cancel, driving $\tilde r_{\mathrm{ss}}\to0$ regardless of $F_{\mathrm{ext}}$ -- the $K_m\to\infty$ limit of the impedance equation at DC, not compliance in its literal sense. This is exactly why Section~\ref{sec:force} measures \emph{higher}, not lower, contact force for the proposed controller (1.47 vs.\ 1.14~mN) than the delayed-feedback baseline: eliminating the position offset by construction eliminates the give that a finite-stiffness law would otherwise provide against a viscoelastic wall. Throughout this paper, ``compliant control'' should therefore be read as naming the receding-horizon \emph{architecture} -- impedance-shaped cost, constraint handling, moving-equilibrium tracking -- not a claim that the benchmarked offset-free instantiation renders finite DC stiffness. The variant that preserves literal finite-$K_m$ compliance omits offset-free force cancellation and accepts the resulting $F_{\mathrm{ext}}/K_m$ offset, trading position accuracy for bounded force; Section~\ref{sec:impedance-family} measures that variant and the intermediate scheduled ones, so the family is reported rather than promised. The alternative is the hard force constraint of Section~\ref{sec:force}, which bounds force directly while keeping offset-free position behavior.

Removing the depth offset generally requires more contact force, so the controller trades placement accuracy against a force limit; the current simulations measure this tradeoff, and a validated force constraint is the next implementation step.

\section{Preview-Based Control Architecture}
\label{sec:architecture}
\begin{figure*}[!t]
\centering
\includegraphics[width=0.92\textwidth]{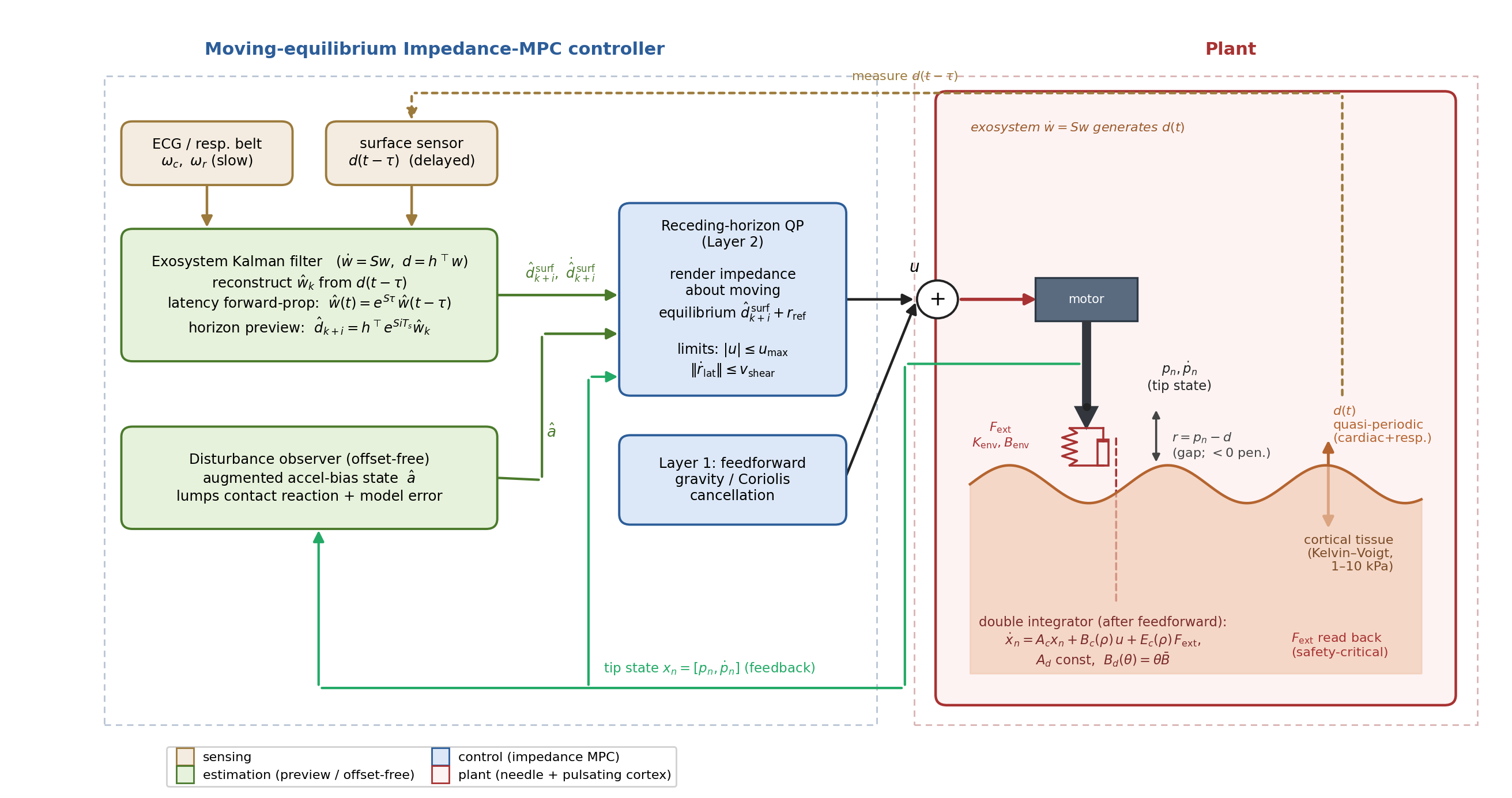}
\caption{Plant and preview-based relative-motion control architecture. Right: the motor-driven insertion axis contacts a pulsating viscoelastic cortical surface through a Kelvin--Voigt contact port. Left: a Kalman predictor estimates the delayed tissue-surface motion and previews it over the horizon; a bias observer compensates persistent contact and model error; the QP regulates relative motion under actuator and shear constraints.}
\label{fig:schematic}
\end{figure*}

Fig.~\ref{fig:schematic} summarizes the plant and controller. The plant is the motor-driven insertion axis, reduced after feedforward compensation to the double integrator of Section~\ref{sec:axis}. Its tip contacts a viscoelastic cortical surface whose quasi-periodic displacement $d(t)$ is generated by the physiological exosystem and coupled through the Kelvin--Voigt contact of Section~\ref{sec:sysmodel}. The controller closes the loop around this moving equilibrium through four elements.

\textbf{Nominal dynamics compensation (Layer 1).} Known robot dynamics -- gravity, Coriolis, task-space inertia -- are cancelled first, exposing the constant-$A_d$ double integrator. Only reliably known terms are cancelled; uncertain or poorly modeled terms are left to the disturbance estimate, which keeps this layer robust by design.

\textbf{Physiological-motion observer.} An exosystem Kalman filter reconstructs $\hat w_k$ from the latency-delayed surface measurement and forward-propagates it by the modeled sensing delay $\tau$, $\hat w(t)=e^{S\tau}\hat w(t-\tau)$, before generating the horizon preview $\hat d^{\mathrm{surf}}_{k+i}$. This improves phase alignment for the modeled periodic component; unmodeled actuator delay is not included in the benchmark.

\textbf{Disturbance estimator (offset-free).} A scalar integrating acceleration-bias state $\hat a$ lumps the contact reaction, friction, and residual model error. Holding it constant over the horizon makes contact tracking offset-free without a force sensor, following standard offset-free and extended-state-observer designs \cite{pannocchia2003,han2009}. Concretely, the observer uses the augmented state $\zeta=[p,v,\hat a]^\top$, position-only measurement $y=p$ ($C=[1,0,0]$), and predict step
\begin{equation}
\begin{aligned}
\zeta_{k+1}^{-}&=A_\zeta\zeta_k+B_\zeta u_k,\\
A_\zeta&=\begin{bmatrix}1&T_s&T_s^2/2\\0&1&T_s\\0&0&1\end{bmatrix},\qquad
B_\zeta=\begin{bmatrix}T_s^2/(2m)\\T_s/m\\0\end{bmatrix},
\end{aligned}
\end{equation}
followed by a fixed-gain Luenberger correction $\zeta_k=\zeta_k^{-}+L\,(y_k-C\zeta_k^{-})$ with hand-tuned gains $L=[l_p,l_v,l_a]^\top=[0.6,\,12.0,\,400.0]^\top$ -- not the solution of a discrete algebraic Riccati equation, so no process or measurement noise covariance is specified. The pair $(A_\zeta,C)$ is observable for $T_s>0$, and the resulting predict-then-correct error matrix $(I-LC)A_\zeta$ has eigenvalues $\{0.40794,\ 0.98993\pm0.02396j\}$, all inside the unit circle at $T_s=1$~ms: one fast real mode and a lightly damped mode near 3.9~Hz ($\approx-9.83\pm24.20j$~rad/s in continuous time) governing how quickly $\hat a$ settles after a step change in contact force. There is no saturation or anti-windup on $\hat a$: a large, fast contact transient such as first tissue contact is tracked by the same linear correction as a slowly drifting bias. That is adequate for the Kelvin--Voigt contact step tested here but is not validated against sharper transients such as vessel puncture or thread release.

\textbf{Constrained MPC (Layer 2).} The receding-horizon QP tracks the moving surface plus the commanded relative-depth profile while limiting actuation and relative velocity; in the three-axis controller it also couples the two lateral velocity components through a polyhedral approximation of a norm bound. Constant $A_d$ means the prediction matrices and QP Hessian are precomputed once.

This architecture splits two roles that delayed feedback alone would otherwise have to perform at once: the harmonic observer accounts for predictable surface motion, while feedback and disturbance estimation correct prediction error and contact-induced bias.

\section{Why Preview Helps Under Delayed Sensing}
\label{sec:whypreview}
Delayed feedback can compensate low-frequency surface motion whenever the sensing rate and loop bandwidth are sufficient, so preview should be read as a way to improve phase margin and disturbance rejection, not as proof that feedback is inherently incapable of meeting the task. For a type-1 loop with gain crossover $\omega_{gc}$, the relative tracking error to a sinusoidal displacement at frequency $\omega$ scales approximately as $(\omega/\omega_{gc})|d(j\omega)|$; a sensing delay adds phase lag $-\omega\tau$, which limits how far the feedback bandwidth can be pushed. The physiological model instead supplies a phase-advanced estimate of the motion without requiring the feedback gain to grow.

This distinction is visible in the 1-DOF benchmark of Section~\ref{sec:sim}. A fair delayed-feedback baseline reaches 18.3\um\ RMS during free-space hover, narrowly meeting the 20\um\ target, while preview reduces the error to 12.2\um. The larger benefit appears after contact, where preview together with the disturbance estimate reduces the steady relative-depth error from 76.0 to 1.1\um.

\section{Controller Design and Stability}
\label{sec:design}

\subsection{Prediction Model}
Stack the LPV double integrator over the horizon, with plant state $x=[p_n,\dot p_n]^\top$ (distinct from the tracking-error state $\xi$ of Proposition~\ref{prop:iss}), decision variables $U=[u_0,\dots,u_{N-1}]^\top$, the offset-free acceleration bias $\hat a$ held constant over the horizon, and the previewed surface from Section~\ref{sec:sysmodel}:
\begin{equation}
\begin{aligned}
&\Xi=\mathcal S_x\,x_0+\mathcal S_u\,U+\mathcal S_a\,\hat a,\\
&\mathcal S_a=\Big[\textstyle\sum_{j} A_d^{\,j}\bar B_a\Big],\ \bar B_a=[T_s^2/2;\,T_s].
\end{aligned}
\end{equation}
The physiological preview enters through the desired state sequence rather than as an unknown disturbance.

\subsection{Cost and the Quadratic Program}
\begin{equation}
\min_{U}\ \sum_{i=1}^{N}\!\Big(q_p(p_{n,i}-p^{\mathrm{des}}_i)^2+q_v(\dot p_{n,i}-\dot p^{\mathrm{des}}_i)^2\Big)+\sum_{i=0}^{N-1}\! r\,u_i^2,
\end{equation}
with moving-equilibrium reference $p^{\mathrm{des}}_i = d^{\mathrm{nom}}_0+\hat d^{\mathrm{surf}}_{k+i}+r_{\mathrm{ref},k+i}$ and ride velocity $\dot p^{\mathrm{des}}_i=\dot{\hat d}^{\mathrm{surf}}_{k+i}$. The weights $(q_p,q_v,r)$ select the rendered impedance ($Q=\mathrm{diag}(K_m,B_m)$ in the equivalence limit of Section~\ref{sec:equiv}). Condensing gives the dense QP
\begin{equation}
\min_U\ \tfrac12 U^\top H U + f^\top U\ \ \text{s.t.}\ \
|u_i|\le u_{\max},\ |\dot p_{n,i}-\dot p^{\mathrm{des}}_i|\le v_{\max},
\end{equation}
\begin{equation}
H=\mathcal S_u^\top \bar Q\,\mathcal S_u+\bar R,\quad
f=\mathcal S_u^\top \bar Q\,(\mathcal S_x x_0+\mathcal S_a\hat a-\Xi^{\mathrm{des}}),
\end{equation}
solved by warm-started OSQP \cite{stellato2020}, with $H$ and its factorization computed once since $A_d$ is constant. The relative-velocity row directly targets the shear surrogate of Section~I. In the 3-axis controller, eight directions $c_j=[\cos\phi_j,\sin\phi_j]^\top$ form an inner octagonal approximation of the lateral-velocity ball. Because these hard rows can become infeasible under degraded sensing, the proposed feasibility-restored formulation adds one shared nonnegative slack per prediction step:
\begin{equation}
\begin{aligned}
\min_{U,s\ge0}\;&\tfrac12U^\top H U+f^\top U+\tfrac{\rho_s}{2}\|s\|_2^2,\\
\mathrm{s.t.}\;&c_j^\top\dot r_{\mathrm{lat},i}\le v_{\mathrm{facet}}+s_i,
\quad j=1,\ldots,8,\quad i=1,\ldots,N.
\end{aligned}
\label{eq:soft-shear}
\end{equation}
The slack is shared across facets so it relaxes the coupled norm geometry rather than one Cartesian axis, and it has no upper bound, so the shear rows cannot make an otherwise input-box-feasible QP infeasible. A quadratic weight $\rho_s=1$ penalizes the physical velocity slack in SI units without the ill-conditioning seen at much larger raw weights; every activation, fallback, measured violation, and placement cost is logged. This restores feasibility, not a hard physical safety guarantee -- the measured plant can still exceed the design budget because of prediction and tracking error. The hard-octagon and no-shear variants also fall back on an infeasible solve rather than aborting, holding the last command and geometrically decaying it toward zero under sustained infeasibility (an unconditional indefinite hold was tested and let the plant diverge); this bounds the failure without restoring shear regulation, which is why the soft formulation above remains necessary.

At each control step, the algorithm:

\begin{algorithmic}[1]
\STATE reads the delayed surface measurement $d(t-\tau)$ and ECG/respiration frequencies;
\STATE updates the exosystem Kalman filter on the delayed sample and propagates $w_{\mathrm{now}}\leftarrow e^{S\tau}w$;
\STATE generates the horizon preview $\hat d^{\mathrm{surf}}_{k+i},\dot{\hat d}^{\mathrm{surf}}_{k+i}$;
\STATE updates the disturbance estimate $\hat a$ from the tip-motion residual;
\STATE builds the moving reference $p^{\mathrm{des}}_i,\,v^{\mathrm{des}}_i$ over the horizon; and
\STATE solves the warm-started QP and applies its first command plus gravity feedforward.
\end{algorithmic}

\subsection{Nominal Impedance Interpretation}
\label{sec:equiv}
With no constraints and $\hat a\equiv0$, the horizon extended to infinity, the controller reduces to LQR for the double integrator. Its Riccati solution produces feedback gains $K=[k_p,k_v]$ and therefore a second-order closed-loop port about the moving equilibrium; in general $k_p,k_v$ are not equal to the diagonal entries of $Q$. This unconstrained limit supplies a reproducible one-parameter initialization: for the normalized plant $\ddot x=u$, the cost ratio $q/r$ sets the natural frequency, and returning to a force-input plant of mass $m$ scales both gains by $m$, so a target nominal stiffness $K_m$ gives
\begin{equation}
q/r=(K_m/m)^2,\qquad K_d=K_m,\qquad B_d=\sqrt{2mK_m},
\end{equation}
with nominal damping ratio $1/\sqrt2$. This is an initialization rule for the unconstrained model, not a uniqueness claim: contact, finite-horizon optimization, active constraints, and disturbance cancellation all alter the realized port, and Section~\ref{sec:impedance-family} measures how far -- the rule is accurate to a factor of three at the benchmarked weights and diverges by two orders of magnitude away from them, because the finite horizon rather than the weight ratio then sets the gain. The derivation, units, and optional frequency-shaped or LPV extensions are collected in Appendix~\ref{app:impedance}. The finite-horizon QP adds preview, offset-free rejection, and shear limits; a hard force row can be added when a validated contact-force model or sensor is available (Section~\ref{sec:force}).

\subsection{Robustness to Reflected-Mass Variation}
\label{sec:robustness}
The post-feedforward plant is uncertain in three ways -- reflected mass $m$, contact stiffness $K_{\mathrm{env}}$, and the exosystem frequencies -- and the controller must tolerate all three. Two mechanisms provide margin, and Section~\ref{sec:robust-sim} measures it.

Because $A_d$ is constant and only $B_d(\theta)=\theta\bar B$ varies, a single quadratic Lyapunov function certifies stability over a mass box by enforcing the synthesis LMI at its two vertices $\theta\in\{\underline\theta,\overline\theta\}$ -- far less conservative than general LPV synthesis, where a parameter-dependent $A(\theta)$ would force a parameter-dependent certificate. For a fixed feedback gain $K$, the closed-loop matrix is $A_{cl}(\theta)=A_d-\theta\bar BK$, and the scalar quadratic form associated with $A_{cl}(\theta)^\top P A_{cl}(\theta)-P$ is convex in $\theta$ for any $P\succ0$; a negative-definite bound at the two endpoints of the mass interval therefore holds throughout it (proved in full as Step~1 of Appendix~\ref{app:proof}). Section~\ref{sec:offline} solves this LMI for $K_N$, the actual first-move gain of the running finite-horizon QP, and reports a verified certificate over $m\in[0.6,1.5]\times$ nominal; the naive symmetric $\pm50\%$ box is \emph{not} certifiable for $K_N$, since it contains a genuinely unstable vertex at $-50\%$, so the margin is real but narrower and asymmetric -- not the box a reader might assume from a $\pm50\%$ mass sweep.

Separately, mass error and contact-stiffness error appear at steady state as a constant force/acceleration bias on the regulated coordinate; the integrating disturbance state $\hat a$ cancels exactly that class, which is why measured tracking is nearly invariant to $+50\%$ mass and $+50\%$ contact stiffness (Section~\ref{sec:robust-sim}), within the mass range where the closed loop itself remains stable.

What neither mechanism covers is broadband, non-harmonic surface motion: the exosystem models only the cardiac/respiratory line spectrum, so a colored-noise or transient component (an ectopic beat, for instance) is neither previewed nor, beyond loop bandwidth, rejected. Section~\ref{sec:robust-sim} quantifies the resulting graceful degradation and motivates an extended-state or unknown-input observer for the broadband residual as the disturbance-side complement to the harmonic exosystem.

For a \emph{guaranteed}, rather than merely measured, bound under bounded estimation error and the parameter box, the shear, force, and actuator constraints would need to be tightened by the worst-case predicted spread (a tube-MPC back-off); the constant-$A_d$ structure keeps that tube cross-section computable offline, but full tube synthesis is future work, and the measured robustness of Section~\ref{sec:robust-sim} is the empirical evidence the construction is meant to certify.

\subsection{Infinite-Horizon Limit and Scope of the Stability Result}
\label{sec:offline}
For inactive constraints, the finite-horizon QP has actual first-move error-feedback gain $K_N=[64.29,\ 1.03]$. Applying the constant-$A_d$/affine-$B_d$ vertex construction of \cite{cao2026interaction} to this gain and the moving preview equilibrium gives a common-$P$ certificate over $m/m_{\mathrm{nom}}\in[0.6,1.5]$, while the naive symmetric $[0.5,1.5]$ box fails and the running QP is itself infeasible at the lower endpoint. The full matrix, conditioning, endpoint spectral radii, raw residuals, and reproduction script are in Appendix~\ref{app:common-p}; scheduled-gain synthesis, embedded timing, memory, and power are not implemented (Appendix~\ref{app:realtime}).

The result is deliberately narrow. Write the tracking-error state as $\xi_k=-[e_k,\dot e_k]^\top=[r_k-r_{\mathrm{ref},k},\ \dot r_k-\dot r_{\mathrm{ref},k}]^\top$.

\noindent\textbf{Assumptions.}
\begin{itemize}
\item \textbf{(A1) Bounded parameter.} $\theta_k=1/m_k\in[\underline\theta,\overline\theta]$ (reflected-inertia box).
\item \textbf{(A2) Bounded estimation/preview error.} The exosystem-KF and frequency-tracker error satisfies $\|\hat w_k-w_k\|\le\varepsilon_w$, the offset-free acceleration-bias error satisfies $\|\hat a_k-a_k\|\le\varepsilon_a$, and the residual broadband disturbance satisfies $\|w^{\mathrm{bb}}_k\|\le\bar w$.
\item \textbf{(A3) Vertex certificate.} There exist $P\succ0$, a gain $K$, and $Q_0\succ0$ solving $A_{cl}(\theta)^\top P A_{cl}(\theta)-P\preceq-Q_0$ at both vertices $\theta\in\{\underline\theta,\overline\theta\}$, with $A_{cl}(\theta):=A_d-B_d(\theta)K$ and a positive-definite stage cost $\ell(\xi,u)\ge\alpha_\ell\|\xi\|^2$. Section~\ref{sec:offline} verifies this is not vacuous by taking $K=K_N$, the actual first-move gain of the running finite-horizon QP, and computing a genuine $(P,Q_0)$ for it over an explicit mass box.
\end{itemize}

\begin{proposition}[Constraint-inactive closed loop]
\label{prop:iss}
Suppose the inverse effective mass lies in the bounded interval of (A1), the preview and bias-estimation errors satisfy (A2), and $P\succ0$, $K$ satisfy the endpoint inequalities of (A3). Then the constraint-inactive tracking-error dynamics $\xi_{k+1}=A_{cl}(\theta_k)\xi_k+G\delta_k$ under $u_k=-K\xi_k$ are input-to-state stable with respect to the bounded preview, bias-estimation, and broadband-disturbance errors: there exist $\beta\in\mathcal{KL}$, $\gamma\in\mathcal K$ with
\begin{equation}
\|\xi_k\|\le \beta(\|\xi_0\|,k)+\gamma\!\big(\max(\varepsilon_w,\varepsilon_a,\bar w)\big),\qquad \forall k\ge0.
\end{equation}
Since $\|\xi_k\|=\|[e_k,\dot e_k]\|$, this bound applies equally to the regulated error. In the nominal limit $\varepsilon_w,\varepsilon_a,\bar w\to0$, $\xi_k\to0$, i.e.\ $e_k\to0$: offset-free tracking.
\end{proposition}
The proof has two steps -- endpoint certification gives a common contraction over the whole mass interval (Step~1), and the bounded residual then enters the Lyapunov inequality as an additive input (Step~2) -- and is given in full in Appendix~\ref{app:proof}.

This proposition is a statement about the vertex-feedback law $u=-K\xi$, not about the constrained receding-horizon QP that the MuJoCo benchmarks actually run.
\begin{remark}[Scope relative to the actively-constrained controller]
\label{rem:constrained-gap}
When the horizon's constraints are inactive at every stage, the error-feedback component of the QP minimizer coincides with $u_k=-K_N\xi_k$ after subtracting the nominal preview/feedforward affine term, so Proposition~\ref{prop:iss} certifies that component. When constraints are \emph{active}, the standard ISS-MPC lifting \cite{mayne2005,limon2009} additionally requires a terminal set $\Omega=\{\xi:\xi^\top P\xi\le c_\Omega\}$ that is constraint-admissible and \emph{robustly}, not merely nominally, invariant under the perturbed transition, i.e.\ $A_{cl}(\theta)\Omega\oplus G\Delta\subseteq\Omega$, together with recursive feasibility of the resulting terminal-set QP. \textbf{Neither is constructed in this paper.} A nominally invariant $\Omega$ is not sufficient on its own: it need not absorb the bounded perturbation $G\delta_k$, so a single perturbed step can leave $\Omega$, and the one-step tail-extension argument used to lift unconstrained ISS to constrained-MPC ISS \cite{mayne2005,limon2009} does not go through. Section~\ref{sec:force} makes this concrete: adding one additional hard force-constraint row to the same QP structure produces solver infeasibility on roughly two-thirds of control ticks at the tested caps -- exactly the failure mode a robustly invariant terminal set and constraint tightening (Section~\ref{sec:robustness}) are meant to rule out. Proposition~\ref{prop:iss} should be read as proved for the constraint-inactive feedback law and as the open target, not a closed gap, for the actively-constrained terminal-set controller. Section~\ref{sec:activity} measures how wide that gap is in practice, and Appendix~\ref{app:terminal} carries the terminal-set construction as far as it goes and reports the number at which it stops.
\end{remark}

\section{Simulation Study}
\label{sec:sim}

\subsection{Setup}
The controller is evaluated in MuJoCo~3.8 (Appendix~\ref{app:reproducibility}). A 1-DOF insertion axis (a motor-actuated needle slide joint) contacts a viscoelastic cortical surface modeled as a second slide joint driven by a stiff, gravity-compensated position servo to a prescribed pulsation $d(t)$. Contact is MuJoCo's soft (Kelvin--Voigt-like) \texttt{solref}/\texttt{solimp} contact; its measured quasi-static effective stiffness ($\sim$3~N/m at the benchmark's contact-patch scale) is not separately calibrated to match the $K_{\mathrm{env}}\approx16$~N/m figure used by the force-constraint model of Appendix~\ref{sec:force} (grounded physiologically there, not fit to this plant); the normal contact force is read back from the simulator directly and is the safety-critical quantity regardless of that mismatch. After gravity feedforward, the needle reduces to the double integrator of Section~\ref{sec:axis}.

\subsection{Cortical-Motion Amplitudes and the Two Benchmark Regimes}
\label{sec:amplitudes}
The pulsation amplitude is the benchmark input a reviewer should question first, because the controller's advantage over a laboratory-frame or delayed-feedback baseline is, by construction, proportional to how much the tissue actually moves. Table~\ref{tab:amplitudes} collects the direct cortical-surface measurements we could find and places both benchmark regimes against them.

\begin{table}[!t]
\renewcommand{\arraystretch}{1.25}
\caption{Measured Cortical-Surface Motion and the Two Benchmark Regimes. Amplitudes Are Single-Sided Unless Noted}
\label{tab:amplitudes}
\centering
\footnotesize
\begin{tabular}{p{2.35cm}p{1.7cm}C{1.5cm}C{1.35cm}}
\toprule
Source / preparation & Method & Normal (\textmu m) & Lateral (\textmu m) \\
\midrule
Rat somatosensory cortex \cite{gilletti2006} & DVRT, surface & 10--30 (resp.), 2--4 (vasc.) & not reported \\
Marmoset, open craniotomy \cite{nomura2024} & imaging & 29.9--39.9 (cardiac) & 1.6--13.0 (cardiac) \\
Human, post-craniotomy \cite{faria2014} & stereo vision & \multicolumn{2}{c}{$<800$ total, not decomposed} \\
\midrule
\textbf{Stress envelope} (this paper) & --- & 200 $+$ 60 $+$ 300 $=$ 560 total & 100--150 \\
\textbf{Measured regime} (Section~\ref{sec:measured-amp}) & --- & 35 $+$ 10.5 $+$ 20 $=$ 65.5 total & 6.7--10 \\
\bottomrule
\end{tabular}
\end{table}

Two things follow. First, the stress envelope is not an arbitrary inflation: its total axial amplitude, 560\um, sits below the ${<}0.8$~mm cortical-surface motion measured by stereo vision immediately after human craniotomy \cite{faria2014}, so it is an approximately human-scale upper bound rather than a value above every published measurement. That figure carries its own caveat and we do not lean on it harder than it can bear: \cite{faria2014} reports the motion of a tracked feature cloud's centre of mass with a system accuracy of 0.5~mm, and does not separate normal from lateral components, so it bounds cortical motion rather than specifying it. Second, the small-animal measurements are an order of magnitude smaller, and the lateral slip figure used in the 3-DOF study, 100--150\um, exceeds every measured cortical lateral value we found by roughly 8--90$\times$; no cortex-specific lateral-slip measurement appears to exist.

Rather than defend one amplitude, we run both regimes. The stress envelope is the headline benchmark throughout; Section~\ref{sec:measured-amp} re-runs the same controllers, unchanged, at the small-animal measured amplitudes of Table~\ref{tab:amplitudes} and reports which conclusions survive and which do not. A species- and procedure-matched cortical-motion measurement for a thread-insertion robot remains a prerequisite for a deployment-grade parameterization.

\subsection{Benchmark Parameters}
The stress-envelope motion parameters are:
\begin{itemize}
\item Cardiac freq./ampl.\ ($f_c,A_c$): 1.2~Hz, 200\um.
\item Cardiac 2nd harmonic ($A_{2c}$): 60\um\ @ 2.4~Hz.
\item Resp.\ freq./ampl.\ ($f_r,A_r$): 0.25~Hz, 300\um.
\item Lateral slip (3-DOF, Section~\ref{sec:multidof}): 100--150\um.
\end{itemize}

\textbf{Engineering/design parameters:}
\begin{itemize}
\item Sensor latency ($\tau$): 15~ms -- OCT/vision loop \cite{ginhoux2005}.
\item Control rate ($1/T_s$): 1~kHz.
\item Effective tip mass ($m$): 1~g -- reduced axis inertia.
\item Tissue stiffness ($K_{\mathrm{env}}$): $\sim$16~N/m -- cortex $\sim$1--10~kPa \cite{budday2017}.
\item Placement / shear spec: $|e|<20$\um, $|\dot e|<2$~mm/s -- low-trauma target.
\end{itemize}

The task: a smooth cosine approach (0--1~s), a free-space hover 0.5~mm above the pulsating surface (1--2~s, the pure motion-compensation test), a ramp into 0.3~mm penetration (2--2.5~s), and a contact hold (2.5--8~s, where tracking, shear, and contact force are measured). The cortical surface pulsates throughout.

Three controllers are compared, all sharing the same gravity feedforward: \textbf{lab-frame PD}, which tracks a fixed setpoint and ignores $d(t)$; \textbf{feedback impedance}, which tracks the delayed measured surface without physiological-motion prediction; and \textbf{preview-based relative-motion control (proposed)}, which combines Kalman prediction with latency forward-propagation, the acceleration-bias observer, and the condensed OSQP QP with shear and actuator limits.

Only the second of these is a fair competitor. Lab-frame PD is included because it makes the paper's premise concrete -- it is what a controller that ignores tissue motion measures, and its error is essentially the pulsation amplitude -- not because beating it is evidence of anything; a reader should read it as an axis label rather than as a baseline. The comparison that carries weight is delayed-feedback impedance, which has the same sensing and the same actuator and is a controller someone would actually build, and, more informative still, the $2\times2$ ablation of Section~\ref{sec:ablation}, which changes one mechanism at a time inside a fixed QP architecture. We did not reimplement an external published method: the closest candidates target different organs, plants, and objectives (Table~\ref{tab:compare}), and a reimplementation retuned by us on our own benchmark would carry less information than the internal ablation, not more.

\subsection{One-Degree-of-Freedom Results}
\begin{table*}[!t]
\renewcommand{\arraystretch}{1.3}
\caption{MuJoCo Benchmark (Measured). Free-Space Window $=$ 1--2~s Hover; Contact Window $=$ 3--8~s Hold}
\label{tab:table1}
\centering
\footnotesize
\begin{tabular}{p{3cm}*{6}{C{1.9cm}}}
\toprule
Controller & RMS $e$ free (\textmu m) & RMS $e$ contact (\textmu m) & peak $|\dot e|$ free (mm/s) & peak $|\dot e|$ contact (mm/s) & peak $F$ contact (mN) & $F$ ripple (mN) \\
\midrule
Lab-frame PD & 314.8 & 232.9 & 2.25 & 2.47 & 2.00 & 2.00 \\
Feedback impedance (no preview) & 18.3 & 76.0 & 1.04 & 0.32 & 1.14 & 1.10 \\
\textbf{Preview relative-motion control (proposed)} & \textbf{12.2} & \textbf{1.1} & \textbf{0.67} & \textbf{0.25} & 1.47 & 1.26 \\
\bottomrule
\end{tabular}
\end{table*}

Lab-frame tracking leaves nearly the full pulsation amplitude in the relative coordinate (314.8\um\ RMS during hover) -- the premise of the paper, stated numerically, rather than a contest. Against the fair baseline the free-space margin is modest: delayed feedback, using a filtered surface-velocity estimate, already reaches 18.3\um\ and meets the 20\um\ specification, and preview improves that to \textbf{12.2\um}. The difference is larger after contact: feedback retains a 76.0\um\ offset under the persistent tissue load, while the acceleration-bias state drives the proposed controller to \textbf{1.1\um} without a force sensor. Measured relative velocity -- from MuJoCo's actual tissue-body velocity, not the prescribed waveform derivative -- is lowest for the proposed controller in both windows (0.25 vs.\ 0.32 vs.\ 2.47~mm/s in contact). Solver timing is reported separately from physical performance because it depends on the host and solver build; full deployed-cycle timing must also include sensing, estimation, and actuator-loop overhead.

\begin{figure}[!t]
\centering
\includegraphics[width=\columnwidth]{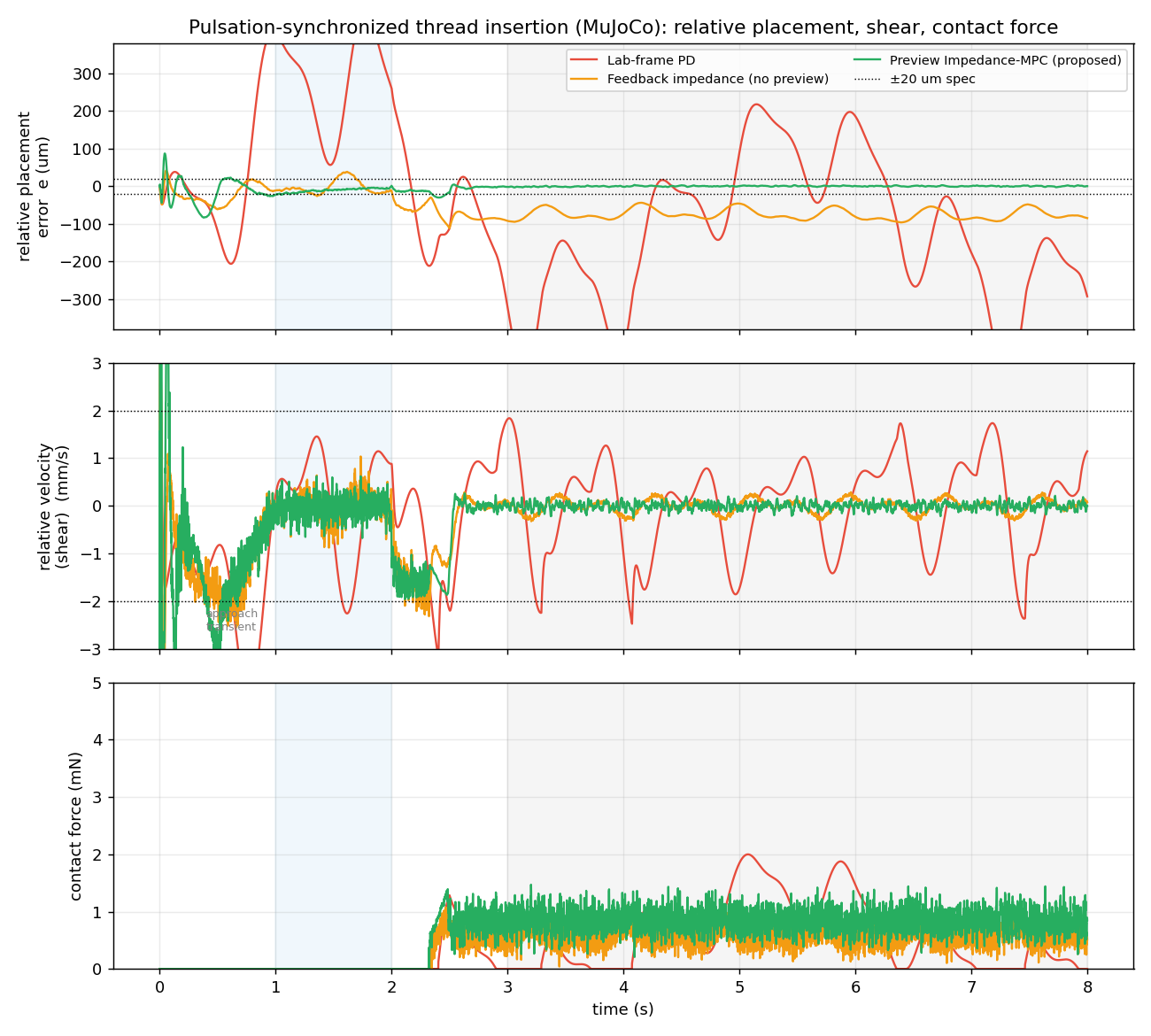}
\caption{1-DOF MuJoCo benchmark (Table~\ref{tab:table1}): relative placement $e$ (top, $\pm$20\um\ spec band), relative velocity (middle), and contact force (bottom) for lab-frame PD, delayed-feedback impedance, and the proposed preview-based controller across the approach, free-space, and contact phases.}
\label{fig:1dof}
\end{figure}

\subsection{Which Mechanism Does the Work: Preview $\times$ Offset-Free}
\label{sec:ablation}
Table~\ref{tab:table1} compares three structurally different controllers, only one of which is a fair competitor; it also reports a single seeded run, so it shows the combined benefit but not each mechanism's individual contribution. This is the more informative comparison, and it is the one a reader should weigh: preview and offset-free are isolated within one fixed QP architecture -- same cost, constraints, and solver, toggling only $\hat a\equiv0$ vs.\ the Luenberger observer and zero-order-hold vs.\ horizon-propagated reference -- and repeats every case over $N=30$ random seeds.

\begin{table}[!t]
\renewcommand{\arraystretch}{1.3}
\caption{2$\times$2 Ablation (Preview $\times$ Offset-Free), Same QP Architecture. Monte Carlo Mean~$\pm$~Std, $N=30$ Seeds}
\label{tab:ablation}
\centering
\scriptsize
\setlength{\tabcolsep}{3pt}
\begin{tabular}{p{1.6cm}*{4}{C{1.28cm}}}
\toprule
Variant & RMS $e$ free (\um) & RMS $e$ cont.\ (\um) & peak $|\dot e|$ cont.\ (mm/s) & peak $F$ cont.\ (mN) \\
\midrule
Neither & $21.1\pm0.2$ & $29.2\pm0.1$ & $1.31\pm0.02$ & $1.49\pm0.06$ \\
Preview only & $11.4\pm0.3$ & $12.7\pm0.1$ & $0.29\pm0.03$ & $1.50\pm0.06$ \\
Offset-free only & $21.1\pm0.2$ & $30.9\pm0.2$ & $1.59\pm0.02$ & $1.56\pm0.06$ \\
\shortstack[l]{\textbf{Full}\\\textbf{(proposed)}} & $\mathbf{11.4\pm0.3}$ & $\mathbf{1.9\pm0.1}$ & $0.29\pm0.03$ & $1.55\pm0.06$ \\
\bottomrule
\end{tabular}
\end{table}

This decomposition changed after the capsule-radius/TIP\_OFFSET correction of Section~\ref{sec:sim}. Preview reduces free-space error (21.1$\to$11.4\um, $\sim$46\%) by anticipating the periodic surface motion ahead of the sensing latency, and does so whether or not offset-free is active (11.4\um\ in both the ``preview only'' and ``full'' rows) -- free-space hover has no persistent contact disturbance for the bias observer to cancel, so offset-free contributes nothing there, as expected. Preview alone now also supplies most of the standalone contact-tracking improvement (29.2$\to$12.7\um, $\sim$2.3$\times$), which was not the case pre-fix, when preview alone was the \emph{worse} standalone mechanism (88.9\um). Offset-free alone, without preview, no longer demonstrates a clear standalone benefit against the corrected (roughly halved) contact-force disturbance: neither (29.2\um) and offset-free-only (30.9\um) are statistically indistinguishable, a reversal of the pre-fix reading where offset-free alone had been the dominant standalone contributor (147.4$\to$30.3\um). We keep the shared observer gain rather than retune it for this one cell: a faster disturbance-observer gain recovers offset-free's standalone benefit here but regresses free-space tracking under tip-mass mismatch elsewhere. The combination remains excellent and is where offset-free's value is actually realized: full contact RMS (1.9\um) is more than $16\times$ better than offset-free alone (30.9\um) and more than $6\times$ better than preview alone (12.7\um) -- offset-free's contribution is concentrated in combination with preview, not as an independently additive term. The Monte Carlo standard deviations are small relative to the between-variant differences that remain real ($\pm0.1$--$0.3$\um\ on the preview and full effects), so those rankings are not an artifact of the single seed used elsewhere in this paper; the offset-free-alone-vs-neither comparison specifically is the one cell not resolved beyond noise.

The original $N=30$ hard-octagon audit predates the corrected logging, preview indexing, and fallback law and is therefore not used as evidence here. The corrected nominal single-seed comparison is reported in Table~\ref{tab:table3}, while the final hard/soft comparison is the matched, checkpointed 10-seed experiment of Tables~\ref{tab:meas-noise} and~\ref{tab:noise-curve}.

\subsection{Placement--Force Tradeoff}
Offset-free tracking increases contact force: the proposed controller reaches the commanded penetration and produces a 1.47~mN peak force, against approximately 1.14~mN for delayed feedback -- whose lower force comes with a 76.0\um\ depth error, so the comparison is a physical tradeoff, not an unqualified safety advantage. A quasi-static penetration row does not reliably cap the Kelvin--Voigt transient force: its hard form is frequently infeasible, while its slack-relaxed form restores feasibility but leaves a 2.5--3~mN measured-versus-predicted gap. A deployed controller should represent the tradeoff explicitly: with a validated force measurement or estimate, a constraint of the form $\hat F_{\mathrm{ext}}\le F_{\max}$ can cap the commanded penetration, and the controller must then accept a bounded depth error when that constraint is active. Appendix~\ref{sec:force} sweeps this tradeoff in full.

One result from that sweep belongs here rather than in an appendix, because it changes what the tradeoff is. The force-constrained rows above cap a \emph{model-predicted} quasi-static force, and that model is the weak link: with a $F_{\max}=2.0$~mN soft row, contact-window RMS placement error is 175.7\um\ and the realized peak force overshoots the cap by 1.70~mN. Feeding the row a simulated force sensor instead -- \texttt{mj\_contactForce} with $\sigma=0.02$~mN of Gaussian noise, used to correct the bound by the gap between sensed and model-predicted force, held constant over the horizon, which is the same offset-free construction the position channel already uses -- moves both quantities at once: peak force comes in \emph{below} the cap ($-0.59$~mN) and contact RMS placement error collapses to 2.1\um, verified stable across ten seeds (Appendix~\ref{sec:realism}). In other words, most of the apparent placement-versus-force tradeoff under a force cap is not physics but the quasi-static contact model's blind spot; correcting it with a real measurement recovers both. Two qualifications keep this honest: it is a simulated sensor with a noise model only, and the benefit does not carry over to the fully combined hardware-realism condition (257.3\um), which that appendix does not explain. Force-sensor feedback therefore remains necessary, and is not yet shown to be sufficient.

\subsection{The Impedance Family and the Stiff-to-Compliant Schedule}
\label{sec:impedance-family}
Section~\ref{sec:impedance} reads motion compensation as impedance control about the moving equilibrium and states that the benchmarked controller sits at the rigid ($K_m\to\infty$ at DC) end of that family. A framing that promises a family should benchmark more than one point of it, so this section sweeps the impedance-shaping weights and schedules them, with everything else -- preview, horizon, constraints, plant, task -- held fixed (\texttt{thread\_insertion\_impedance\_family.py}).

\begin{table}[!t]
\renewcommand{\arraystretch}{1.2}
\caption{Impedance Family (Measured). Input Weight $r$ Swept at $q_p=1$; Contact-Window RMS and Peak Force with the Offset-Free State On / Off}
\label{tab:impedance}
\centering
\scriptsize
\setlength{\tabcolsep}{4pt}
\begin{tabular}{ccccc}
\toprule
$r$ & nom.\ $K_m$ & real.\ $k_p$ & RMS $e$ cont.\ (\textmu m) & peak $F$ (mN) \\
 & (N/m) & (N/m) & on / off & on / off \\
\midrule
$2\times10^{-9}$ & 22.36 & 64.29 & 1.8 / 12.7 & 1.48 / 1.43 \\
$2\times10^{-7}$ & 2.24 & 64.23 & 1.8 / 12.7 & 1.48 / 1.43 \\
$2\times10^{-5}$ & 0.224 & 58.86 & 1.8 / 13.8 & 1.44 / 1.39 \\
$2\times10^{-4}$ & 0.071 & 39.56 & 1.9 / 19.8 & 1.28 / 1.20 \\
$2\times10^{-3}$ & 0.022 & 16.99 & 2.7 / 41.6 & 1.07 / 0.90 \\
$2\times10^{-2}$ & 0.0071 & 5.40 & 7.9 / 101.5 & 1.01 / 0.59 \\
$2\times10^{-1}$ & 0.0022 & 1.21 & 68.8 / 202.4 & 0.67 / 0.27 \\
\bottomrule
\end{tabular}
\end{table}

Two findings, one of which corrects the framing. First, the infinite-horizon initialization rule of Appendix~\ref{app:impedance} is not predictive away from its tuning point. $K_N$ maps position error directly to commanded force, so its first entry \emph{is} the realized DC stiffness and is directly comparable to the nominal $K_m$: at the benchmarked weights the two agree to within a factor of three (64.29 vs.\ 22.36~N/m), but a $10^{4}$ change in $r$ moves the nominal figure by $10^{2}$ and the realized gain by only 1.1, because below $r\approx2\times10^{-4}$ the $N=25$ horizon, not the weight ratio, sets the gain. ``Rendered stiffness'' is therefore indexed by the realized first-move gain, not by the cost weights; Appendix~\ref{app:impedance} should be read as an initialization at one operating point, which is what it claims, and not as a design map.

Second, once $r$ is large enough to reach the realized gain, the family behaves as impedance control should, and the schedule the framing describes is worth having:

\begin{table}[!t]
\renewcommand{\arraystretch}{1.2}
\caption{Stiff-to-Compliant Schedule (Measured). Stiff Weights in Free Space, Switching at the Contact Transition ($t=2.5$~s)}
\label{tab:schedule}
\centering
\scriptsize
\setlength{\tabcolsep}{4pt}
\begin{tabular}{lcccc}
\toprule
Contact-phase wts. & off-free & RMS $e$ free & RMS $e$ cont. & peak $F$ \\
 & & (\textmu m) & (\textmu m) & (mN) \\
\midrule
none (benchmarked, all stiff) & on & 12.0 & 1.8 & 1.48 \\
$r=2\times10^{-3}$ & on & 12.0 & 2.7 & \textbf{1.07} \\
$r=2\times10^{-2}$ & on & 12.0 & 8.0 & 1.01 \\
$r=2\times10^{-3}$ & off & 12.0 & 41.6 & 0.90 \\
$r=2\times10^{-2}$ & off & 12.0 & 101.5 & \textbf{0.59} \\
\bottomrule
\end{tabular}
\end{table}

Scheduling stiff weights during free-space approach and compliant weights after contact leaves free-space tracking untouched at 12.0\um\ -- as it must, since the switch happens at $t=2.5$~s -- and buys a real reduction in peak contact force. The most useful operating point is not the one the paper benchmarks: scheduling to $r=2\times10^{-3}$ with the offset-free state kept on reaches 1.07~mN peak force at 2.7\um\ contact RMS, against 1.48~mN at 1.8\um\ for the all-stiff benchmarked controller -- a 28\% force reduction for 0.9\um\ of placement error. Turning the offset-free state off as well produces the literal finite-$K_m$ behavior, with the expected steady deflection: 0.59~mN at 101.5\um.

This settles what ``compliant control'' names in this paper. The dominant knob is the offset-free bias state, not the cost weights: at every weight setting in Table~\ref{tab:impedance} the on/off difference is larger than the effect of a decade of $r$. The cost weights supply a genuine, monotone placement--force trade on top of it once they are large enough to matter. Both are now measured rather than described; what remains unbenchmarked, and is stated as an extension rather than a result, is frequency-shaped or LPV-scheduled impedance (Appendix~\ref{app:impedance}) and any claim that a particular rendered port is the clinically right one, which needs a force or energy limit this paper does not have.

\subsection{Model Mismatch and Nonperiodic Motion}
\label{sec:robust-sim}
The proposed controller is re-run with the plant perturbed while the controller keeps its nominal assumptions ($m=1$~g, $f_c=1.2$~Hz, $\tau=15$~ms), so every case below is a genuine model mismatch the controller must reject. Each entry reports RMS relative-placement error in free space and in contact, peak contact relative velocity (shear), and peak contact force, all measured:
\begin{itemize}
\item \textbf{Nominal:} 12.2 / 1.1\um\ RMS $e$; 0.25~mm/s shear; 1.47~mN peak $F$.
\item \textbf{Tip mass $+50\%$:} 12.2 / 1.1\um; 0.20~mm/s; 1.93~mN.
\item \textbf{Tip mass $-40\%$ (verified box edge):} 12.2 / 1.2\um; 0.34~mm/s; 1.09~mN.
\item \textbf{Tip mass $-50\%$:} QP infeasible at $t=0.14$~s (untested, unstable vertex).
\item \textbf{Cardiac freq.\ $+8\%$ detune:} 12.4 / 11.0\um; 0.43~mm/s; 1.56~mN.
\item \textbf{Latency 20~ms (assumes 15):} 14.6 / 6.7\um; 0.31~mm/s; 1.53~mN.
\item \textbf{Contact $+50\%$ stiffer:} 12.2 / 1.1\um; 0.24~mm/s; 2.12~mN.
\item \textbf{Non-harmonic 40\,\um\ $+$ ectopic:} 37.3 / 42.2\um; 8.49~mm/s; 5.71~mN.
\item \textbf{All combined (worst case):} 44.3 / 50.4\um; 8.87~mm/s; 7.28~mN.
\end{itemize}

Parametric mismatch is absorbed within the verified stability range and fails genuinely outside it: a $+50\%$ reflected-mass error and a $+50\%$ contact-stiffness error leave tracking essentially unchanged (contact RMS $\approx1.1$\um), since the offset-free state cancels the induced constant bias and the fixed gain remains stable -- matching the two-vertex certificate verified for the actual running gain over $m\in[0.6,1.5]\times$ nominal (Section~\ref{sec:offline}). The $-40\%$ case, the verified box's other edge, still tracks well (1.2\um, 1.09~mN); only the peak force rises with a stiffer or heavier plant, motivating the force-constrained variant of Appendix~\ref{sec:force}. A $-50\%$ case, outside the verified box, is a genuine failure: the closed loop is linearly unstable there, and the running QP reports \texttt{OSQP primal infeasible} 0.14~s into the simulation.

Preview detuning degrades gracefully: an 8\% cardiac-frequency error or a 33\% latency error keep free-space tracking inside spec (12.4 / 14.6\um) and contact tracking in the single-digit-to-low-tens of microns, since the integrating disturbance state recovers most of what the detuned preview loses. Broadband motion is the real limit: adding a non-harmonic 40\um-RMS component and an ectopic-beat transient pushes placement to $\sim$37--42\um\ and measured contact relative velocity to 8.49~mm/s, since the harmonic exosystem cannot represent that content and feedback cannot reject it beyond loop bandwidth.

\subsection{Three-Degree-of-Freedom Shear Regulation and Feasibility Recovery}
\label{sec:multidof}
This is the section the paper's main claim rests on. Axial compensation alone does not address lateral tissue motion: heartbeat and respiration drive 3-D local expansion and lateral slip of the cortical surface, and lateral relative velocity is a controllable mechanical surrogate for shear (Section~\ref{sec:surrogate}) that an axial-only controller structurally misses. The resulting safety requirement couples the control axes and cannot be expressed as a bank of independent SISO loops -- and, as the degraded-sensing results below show, cannot usefully be expressed as a hard constraint either.

The 3-DOF benchmark adds lateral $x,y$ motion, a 3-D pulsating surface, and Coulomb friction, and compares lab-frame PD (3-axis), axial-only preview (full preview and offset-free control of $z$, with $x,y$ held by lab-frame PD), and coupled 3-DOF preview control, which adds the eight-facet coupled lateral-shear constraint of~\eqref{eq:soft-shear} in both its original hard-octagon form and the proposed soft-slack form.

The lateral-slip amplitude is the least grounded input in the paper (Section~\ref{sec:amplitudes}), so it is swept rather than asserted (Table~\ref{tab:table4b}), and the whole 3-DOF study is repeated at measured amplitudes in Section~\ref{sec:measured-amp}.

\begin{table*}[!t]
\renewcommand{\arraystretch}{1.2}
\caption{3-DOF MuJoCo Result at 5\um\ RMS Per-Axis Sensor Noise (Measured). Shear Budget $0.80$~mm/s}
\label{tab:table3}
\centering
\footnotesize
\begin{tabular}{p{4.2cm}*{6}{C{1.65cm}}}
\toprule
Controller & lat. place. RMS (\textmu m) & ax. place. RMS (\textmu m) & peak lat. shear, free (mm/s) & peak lat. shear, cont. (mm/s) & peak normal $F$ (mN) & peak lat. friction (mN) \\
\midrule
Lab-frame PD (3-axis) & 134.3 & 246.2 & 1.33 & 1.36 & 1.02 & 0.13 \\
Axial-only preview MPC & 134.6 & 1.3 & 1.33 & 1.47 & 0.75 & 0.09 \\
Coupled, hard octagon & 1.7 & 1.3 & 0.57 & 0.49 & 0.80 & 0.37 \\
\textbf{Coupled, soft octagon (proposed)} & \textbf{1.7} & \textbf{1.3} & \textbf{0.57} & \textbf{0.49} & 0.80 & 0.37 \\
\bottomrule
\end{tabular}
\end{table*}

At nominal (5\um) sensing, the axial-only controller is excellent axially (1.3\um) but ineffective laterally (135\um, 1.47~mm/s shear); the coupled MPC's tightened facet bounds ($v_{\mathrm{facet}}=0.55\,v_{\mathrm{shear}}\cos(\pi/8)$, backed off to cover observer and plant-tracking error) hold measured shear near budget.

\begin{figure}[!t]
\centering
\includegraphics[width=\columnwidth]{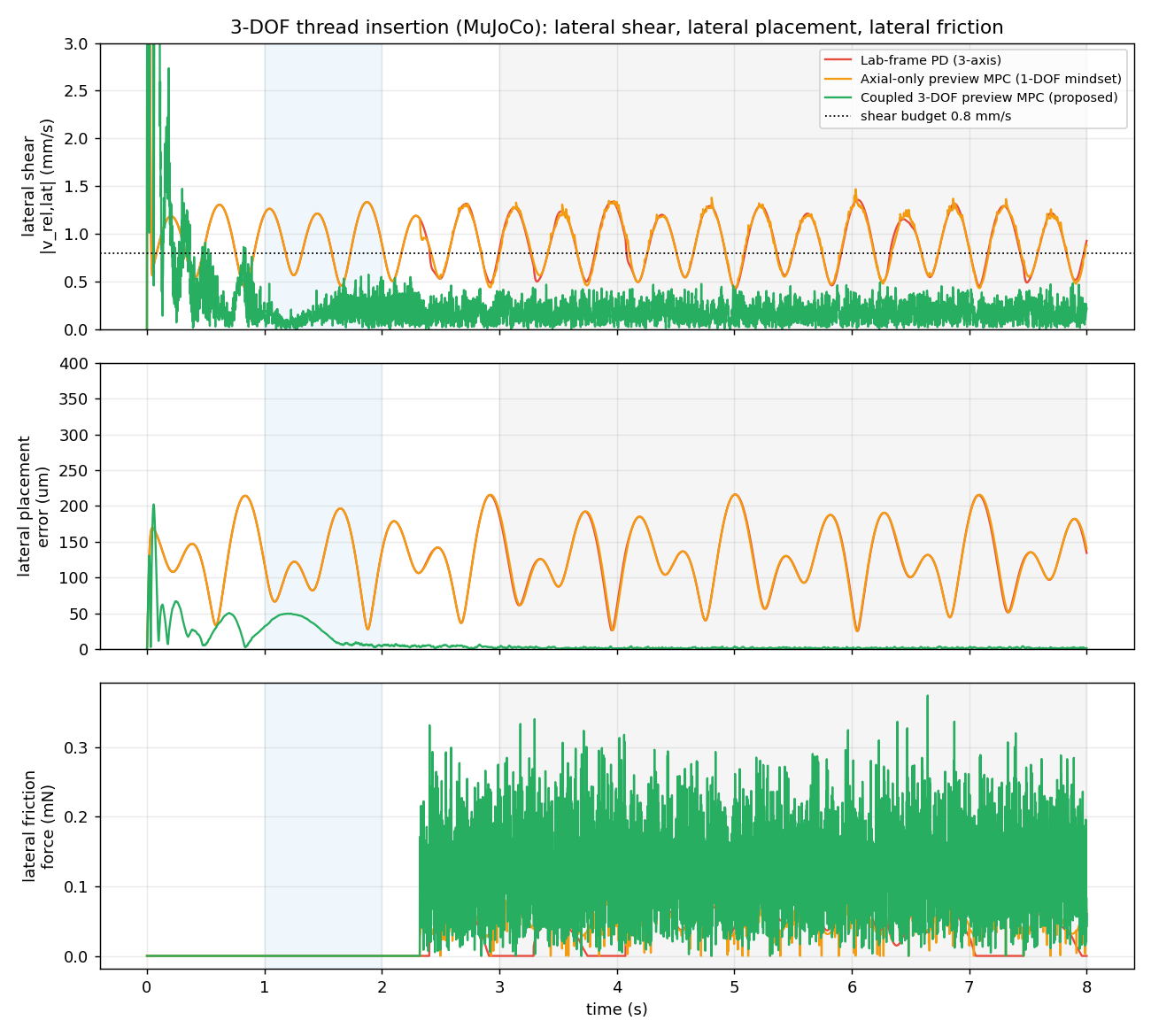}
\caption{3-DOF MuJoCo benchmark (Table~\ref{tab:table3}): lateral and axial placement, shear velocity, and contact force for lab-frame PD, axial-only preview, and coupled preview control.}
\label{fig:3dof}
\end{figure}

\begin{table}[!t]
\renewcommand{\arraystretch}{1.3}
\caption{Lateral-Slip Amplitude Sensitivity (Measured). Peak Contact Lateral Shear (mm/s); Budget $v_{\mathrm{shear}}=0.80$~mm/s}
\label{tab:table4b}
\centering
\begin{tabular}{ccc}
\toprule
Lateral slip $(x,y)$ (\textmu m) & Axial-only MPC & Coupled MPC \\
\midrule
(50, 33) & 0.44 & 0.49 \\
(100, 67) & 0.90 & 0.48 \\
(150, 100) & 1.35 & 0.48 \\
(225, 150) & 2.06 & 0.49 \\
(300, 200) & 2.76 & 0.50 \\
\bottomrule
\end{tabular}
\end{table}

The axial-only shear increases approximately with lateral amplitude; the tightened coupled MPC remains below the 0.80~mm/s measured norm budget throughout this sweep. At the smallest amplitude, active lateral tracking is unnecessary and slightly increases the peak (0.49 versus 0.44~mm/s).

\textbf{Feasibility recovery under degraded sensing.} The explicit constraint's benefit over lateral preview alone is modest at the nominal noise level; what separates the three formulations is what happens when the sensing degrades. At 10\um\ RMS per-axis noise, a cost-only controller -- lateral preview and offset-free tracking, shear penalized in the cost but not constrained -- violates the 0.80~mm/s budget in 10 of 10 seeds. The hard-octagon form fixes that on nine seeds and fails badly on the tenth: its horizon becomes persistently infeasible, and the bounded, geometrically decaying command fallback that keeps the run from aborting still leaves measured shear at 307.9~mm/s, roughly 385$\times$ the budget it was installed to enforce. The soft-slack formulation of~\eqref{eq:soft-shear} completes every seed with 0 of 10 violations and never exhibits that outlier.

The failure mode is worth naming precisely, because it is what motivates the contribution rather than a numerical curiosity. A hard constraint does not degrade: when the tightened rows and the horizon state conflict, the QP returns \texttt{primal infeasible} and the commanded input is produced by whatever fallback the implementation happens to have, which is now the safety controller and was never designed as one. The shared slack removes the failure by construction -- the slack is unbounded above, so the shear rows cannot make an otherwise input-box-feasible QP infeasible -- while keeping the coupled geometry, since one slack serves all eight facets rather than relaxing a single Cartesian axis. What is bought is solvability, not a hard physical guarantee: the measured plant can still exceed the design budget through prediction and tracking error. Appendix~\ref{app:multidof} reports the corrected 10\um\ operating point, including placement error, slack magnitude, violation duration, and solver timing across all three formulations.

\subsection{The Same Benchmark at Measured Cortical Amplitudes}
\label{sec:measured-amp}
Everything above runs at the stress envelope of Table~\ref{tab:amplitudes}. This section re-runs the identical controllers -- same gains, horizon, weights, latency, sensor noise, contact model, task profile, budgets -- at the directly measured small-animal amplitudes of that table, roughly one sixth of the stress envelope axially and one fifteenth laterally (\texttt{thread\_insertion\_measured\_amplitude.py}). The point is not to claim the smaller regime is the right one; it is that the reader should not have to guess which conclusions depend on the amplitude, and three of them do.

\begin{table}[!t]
\renewcommand{\arraystretch}{1.25}
\caption{1-DOF Benchmark at Measured Cortical Amplitudes (Measured). Compare Table~\ref{tab:table1}}
\label{tab:meas1dof}
\centering
\footnotesize
\begin{tabular}{p{2.5cm}*{3}{C{1.35cm}}}
\toprule
Controller & RMS $e$ free (\textmu m) & RMS $e$ contact (\textmu m) & peak $F$ contact (mN) \\
\midrule
Lab-frame PD & 31.8 & 80.4 & 0.74 \\
Feedback impedance & \phantom{0}3.4 & 74.7 & 1.09 \\
\textbf{Proposed} & \textbf{\phantom{0}1.7} & \textbf{\phantom{0}0.9} & 1.47 \\
\bottomrule
\end{tabular}
\end{table}

\textbf{What changes.} Preview's free-space margin is amplitude-dependent, and shrinks to the point of not being needed. At the stress envelope the delayed-feedback baseline reaches 18.3\um\ and only just meets the 20\um\ specification, so preview's 12.2\um\ is the difference between marginal and comfortable; at measured amplitudes that baseline reaches 3.4\um\ and meets the specification with a factor of six to spare. Preview still improves on it (1.7\um), but a reader deciding whether to implement a physiological-motion observer for a small-animal preparation should conclude from Table~\ref{tab:meas1dof} that delayed feedback suffices for free-space hover, and that the case for preview at that scale has to be made on the contact phase or on human-scale motion. Lab-frame PD, correspondingly, stops being catastrophic in free space (31.8 rather than 314.8\um) though it still misses the specification.

\textbf{What does not change.} The offset-free contact result is amplitude-invariant, because the disturbance it cancels is the contact load, not the surface motion: 74.7\um\ for delayed feedback against 0.9\um\ for the proposed controller, a factor of approximately 79, against approximately 67 at the stress envelope. The placement--force trade behaves the same way (1.47 vs.\ 1.09~mN).

\begin{table}[!t]
\renewcommand{\arraystretch}{1.25}
\caption{3-DOF Benchmark at Measured Cortical Amplitudes, 5\um\ RMS Per-Axis Noise (Measured). Compare Table~\ref{tab:table3}}
\label{tab:meas3dof}
\centering
\footnotesize
\begin{tabular}{p{2.7cm}*{3}{C{1.25cm}}}
\toprule
Controller & lat.\ place.\ RMS (\textmu m) & peak lat.\ shear, cont.\ (mm/s) & peak normal $F$ (mN) \\
\midrule
Lab-frame PD (3-axis) & 9.0 & 0.09 & 0.31 \\
Axial-only preview & 9.0 & 0.10 & 0.75 \\
Coupled, hard octagon & 1.6 & 0.49 & 0.80 \\
\textbf{Coupled, soft octagon} & \textbf{1.6} & 0.49 & 0.80 \\
\bottomrule
\end{tabular}
\end{table}

\textbf{The result that reverses.} At measured lateral amplitudes with clean sensing, the coupled shear constraint does not earn its keep, and lateral riding actively costs shear. The axial-only controller leaves 0.10~mm/s of peak contact shear -- nearly an order of magnitude inside the 0.80~mm/s budget -- because a 10\um\ lateral excursion at 1.2~Hz simply does not produce much relative velocity. The coupled controller reduces lateral placement error from 9.0 to 1.6\um\ but \emph{raises} peak shear to 0.49~mm/s, since a tip commanded to ride a 10\um\ lateral signal measured with 5\um\ RMS noise spends much of its effort chasing the noise. Both remain within budget, so this is a genuine trade between lateral placement and lateral shear rather than a failure, but it is not the story Table~\ref{tab:table3} tells, where the same comparison is 135\um\ and 1.47~mm/s against 1.7\um\ and 0.49~mm/s. The (50,\,33)\um\ row of Table~\ref{tab:table4b} was the first sign of this and the measured regime confirms it: \emph{if} cortical lateral slip is of the order the small-animal measurements report, an axial-only controller meets the shear budget on its own and the coupled constraint is insurance rather than necessity.

\begin{table*}[!t]
\renewcommand{\arraystretch}{1.15}
\caption{Targeted Noise--Shear--Feasibility Result at 10\um\ RMS Per-Axis Noise, 10 Seeds, at Measured and Stress-Envelope Amplitudes. Entries Are Mean / Worst Measured Peak Contact Shear (mm/s) and the Number of Seeds with Any $0.80$~mm/s Exceedance}
\label{tab:meas-noise}
\centering
\footnotesize
\begin{tabular}{c*{6}{C{1.9cm}}}
\toprule
\multirow{2}{*}{$\sigma_y$ (\textmu m)} & \multicolumn{2}{c}{no row} & \multicolumn{2}{c}{hard octagon} & \multicolumn{2}{c}{soft octagon} \\
 & measured & stress & measured & stress & measured & stress \\
\midrule
10.0 & 1.194 / 1.420 (10) & 1.195 / 1.419 (10) & 31.34 / 307.4\rlap{$^*$} (1) & 31.39 / 307.9\rlap{$^*$} (1) & 0.659 / 0.695 (0) & 0.662 / 0.698 (0) \\
\bottomrule
\end{tabular}
\end{table*}
$^*$The same seed (index 8) in both regimes: its horizon becomes persistently infeasible for 7998 control ticks, and the geometrically decaying command fallback avoids an abort but not the excursion. Excluding it, the other nine seeds average 0.665 / 0.692~mm/s at measured amplitudes and 0.669 / 0.695~mm/s at the stress envelope.

\textbf{The result that does not move at all.} Table~\ref{tab:meas-noise} places the corrected 10\um\ degraded-sensing study side by side in the two regimes, and the results agree within a few percent for every formulation -- including the hard octagon's catastrophic seed, which is seed~8 in both regimes with approximately 308~mm/s of measured shear. The result is therefore noise-driven rather than motion-amplitude-driven at this operating point. The paper's central claim -- that a coupled shear constraint must be formulated to preserve feasibility, and that the hard form fails exactly where it is needed -- does not rest on the amplitude assumption that is the benchmark's weakest input.

\subsection{How Much of the Benchmark the Stability Result Covers}
\label{sec:activity}
Proposition~\ref{prop:iss} is proved for the constraint-inactive feedback law, and Remark~\ref{rem:constrained-gap} states that this is not the controller the benchmarks run. That gap is usually left qualitative. It need not be: the running QP can report, at each tick, whether its own solution touches a constraint, and the fraction of ticks on which it does is a direct measure of how much of the benchmark the proposition actually covers. A tick is counted as constraint-active if the returned optimizer meets the input box or a relative-velocity (or, in 3-DOF, octagon) row to within $10^{-4}$ of the bound; the audit is passive and never feeds back into the command. Table~\ref{tab:activity} reports the result.

\begin{table}[!t]
\renewcommand{\arraystretch}{1.25}
\caption{Constraint-Active Fraction of Control Ticks (Measured), Proposed Controller}
\label{tab:activity}
\centering
\footnotesize
\begin{tabular}{lcc}
\toprule
Window & 1-DOF & 3-DOF coupled \\
\midrule
Whole run (0--8~s) & 6.0\% & 22.1\% \\
Approach (0--1~s) & 31.2\% & --- \\
Free-space hover (1--2~s) & \textbf{0.0\%} & 0.0\% \\
Penetration ramp (2--2.5~s) & 33.0\% & --- \\
Contact hold (3--8~s) & \textbf{0.0\%} & 4.4\% \\
\bottomrule
\end{tabular}
\end{table}

For the 1-DOF controller, the constraints are inactive on 94\% of control ticks, and -- the point that matters for reading Table~\ref{tab:table1} -- on 100\% of the ticks inside both measurement windows. All of the activity is the relative-velocity row, none of it the input box, and it is confined to the two commanded transients: the cosine approach from a 1.7~mm initial gap and the penetration ramp, where the reference itself demands more relative velocity than the row allows. Every number reported for the 1-DOF benchmark is therefore measured in a regime where Proposition~\ref{prop:iss}'s hypothesis holds exactly, which is a stronger statement than the proposition being ``about a related controller''.

The 3-DOF coupled controller is the honest counterexample. Its shear rows are active on 22.1\% of ticks overall and 4.4\% of the contact window, because bounding shear is the point of the controller: a constraint that never binds is not doing anything. The theory--experiment gap is thus not uniform across the paper -- it is negligible for the 1-DOF results and real, though bounded, for the 3-DOF shear results, which is precisely where feasibility recovery (Section~\ref{sec:multidof}) rather than a stability certificate is what protects the closed loop today.

\subsection{Cardiac-Rate Drift}
\label{sec:freqadapt}
A harmonic preview model is useful only if its frequency stays aligned with the measured motion, which matters most when surface measurements are sparse: at 1~kHz sensing the observer is corrected almost every step and a stale cardiac frequency is mainly suboptimal, but at 25~Hz the observer must extrapolate between measurements and phase error can accumulate. A realistic 0.4~Hz rate drift over the 15~ms sensing latency alone produces only a $\sim2^\circ$ phase error, so the danger is specifically sparse sensing, not the latency itself.

We add a frequency-adaptive observer -- a band-pass-plus-zero-crossing period tracker, a lightweight stand-in for a SOGI-FLL or an EKF on $\omega_c$ -- whose estimate $\hat\omega_c$ rebuilds $A_d$ and $e^{S\tau}$ analytically each step. The cardiac rate is ramped 1.2$\to$1.6~Hz over $t\in[2,5]$~s during free-space hover, and feedback, fixed-frequency preview, and adaptive-frequency preview are compared at both 25~Hz (sparse) and 1~kHz (dense) surface sensing.

\begin{table}[!t]
\renewcommand{\arraystretch}{1.3}
\caption{Cardiac-Rate Drift (1.2 $\rightarrow$ 1.6~Hz); Free-Space Hover (Measured). Pre-Drift Window 1--2~s; Drifted Window $t\ge5.5$~s. Shear $=$ Peak Relative Velocity}
\label{tab:table4}
\centering
\scriptsize
\begin{tabular}{p{0.85cm}p{1.55cm}*{4}{C{0.95cm}}}
\toprule
Sensing & Controller & pre RMS $e$ (\textmu m) & drift RMS $e$ (\textmu m) & drift peak $e$ (\textmu m) & drift shear (mm/s) \\
\midrule
\textbf{25 Hz} & Feedback (no prev.) & 46.0 & 69.8 & 157.3 & 3.44 \\
(sparse) & Fixed-freq preview & 24.0 & 70.8 & 154.2 & \textbf{6.53} \\
 & \textbf{Adaptive-freq} & 37.6 & \textbf{24.8} & \textbf{72.2} & 3.42 \\
\midrule
\textbf{1 kHz} & Feedback (no prev.) & 23.1 & 37.1 & 76.0 & 1.31 \\
(dense) & Fixed-freq preview & 11.8 & 37.1 & 66.5 & 1.48 \\
 & \textbf{Adaptive-freq} & 19.2 & \textbf{12.7} & \textbf{37.1} & \textbf{1.07} \\
\bottomrule
\end{tabular}
\end{table}

\begin{figure}[!t]
\centering
\includegraphics[width=\columnwidth]{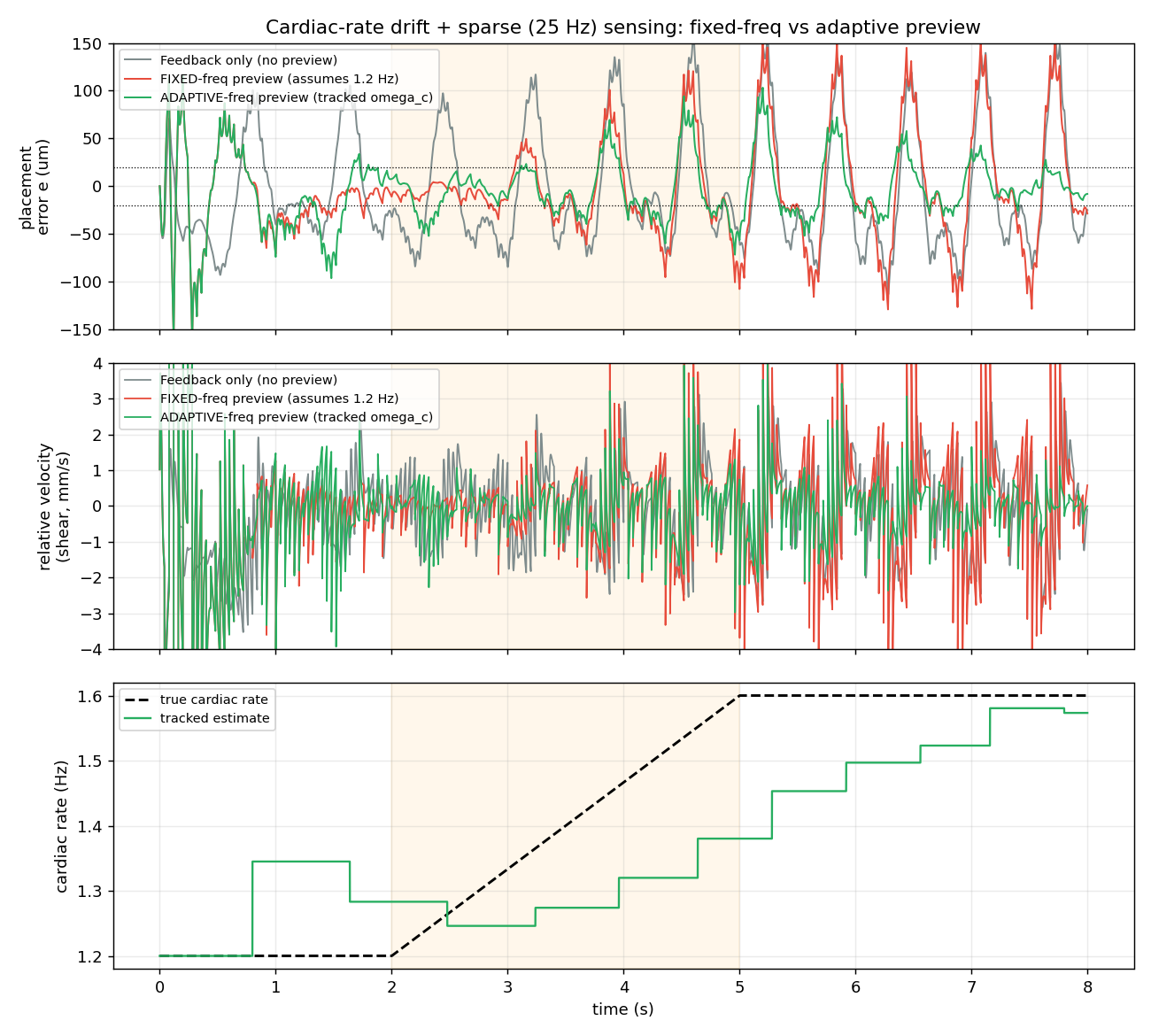}
\caption{Cardiac-rate drift (1.2$\to$1.6~Hz) in free-space hover (Table~\ref{tab:table4}): relative placement and velocity for feedback, fixed-frequency preview, and adaptive-frequency preview.}
\label{fig:freqadapt}
\end{figure}

Fixed-frequency preview can be worse than feedback under sparse sensing: at 25~Hz it produces 6.53~mm/s drift-window peak relative velocity, against 3.44~mm/s for feedback alone. Frequency adaptation is a real improvement, cutting drift-window RMS error to 24.8\um\ at 25~Hz and 12.7\um\ at 1~kHz (relative velocity 3.42 and 1.07~mm/s), but it does not fully restore the nominal design spec at 25~Hz: the 1~kHz case returns within the 2~mm/s design target, the 25~Hz case does not. Dense sensing limits open-loop phase drift -- at 1~kHz the fixed observer is suboptimal rather than catastrophic, though its 1.48~mm/s still exceeds feedback's 1.31~mm/s -- and adaptation has a convergence cost: pre-drift error is higher with the adaptive observer than the fixed one at 25~Hz (37.6 vs.\ 24.0\um). A better-characterized EKF or frequency-locked loop is needed before this estimator is production-ready.

\section{Real-Time Implementation}
\label{sec:realtime}
The 1-DOF QP solves in 40.8~\textmu s mean, 45.8~\textmu s at the 95th percentile, and 660.6~\textmu s at worst; the 3-DOF coupled QP solves in 224.3~\textmu s mean, 239.8~\textmu s at the 95th percentile, and 1057.7~\textmu s at worst. A 1~kHz solver rate is therefore supported for the 1-DOF problem but not yet demonstrated for the full 3-DOF loop.

A more credible implementation separates prediction from fast actuation: the MPC can run at 200--500~Hz to update the motion preview, constraints, and force command, while an inner current or force loop runs at 2--5~kHz, preserving fast actuator regulation without requiring every multivariable QP to finish within 1~ms. The constant state-transition matrix also lets most prediction terms and the QP Hessian be precomputed, so warm starting limits each update to the changing state, preview, disturbance estimate, and constraint bounds. For tighter embedded budgets, an offline scheduled feedback law could replace the QP when constraints are inactive, though its timing, memory use, numerical precision, and power consumption have not been evaluated here. Appendix~\ref{app:realtime} discusses this deployment path further.

\section{Discussion}
The framing result is straightforward: insertion accuracy should be defined relative to the tissue, not the laboratory. Once the reference is written as $d+r_{\mathrm{ref}}$, one controller can follow the pulsating surface during approach and regulate depth after contact.

The transferable technical result is narrower and, we think, more useful. A safety requirement stated as a coupled bound on lateral tip--tissue velocity cannot be met by imposing it as a hard constraint: under degraded sensing the bound conflicts with the horizon state, the QP becomes infeasible, and whatever fallback the implementation chooses is now the safety controller. Section~\ref{sec:multidof} shows what that costs -- one seed in ten at 10\um\ sensing noise ends in a persistently infeasible horizon whose bounded fallback still leaves an approximately 308~mm/s excursion -- and that a shared-slack relaxation of the same rows removes the failure mode without giving up the coupled geometry. This is not specific to neural insertion: any receding-horizon controller enforcing a coupled kinematic safety bound against an estimated moving target faces the same choice, and the empirical answer here is that feasibility restoration matters more than constraint hardness. The same experiment repeated at directly measured cortical amplitudes (Section~\ref{sec:measured-amp}) returns the same ranking, because at this noise level the shear being bounded is noise-driven rather than motion-driven.

Three further points frame the rest. First, accurate depth regulation and low contact force are distinct objectives: offset-free tracking produces more force precisely because the tip reaches the requested penetration. Section~\ref{sec:impedance-family} measures the family that trades between them, and finds a scheduled operating point (1.07~mN at 2.7\um) that dominates the benchmarked one on force at negligible placement cost; a clinically meaningful choice within that family still needs a validated force or energy limit. Second, the harmonic observer is effective for cardiac and respiratory lines; broadband or transient motion -- the ectopic-beat experiment causes the largest loss of tracking and shear performance -- needs a hybrid observer combining harmonic prediction with an unknown-input or broadband residual estimate. Third, the formal stability guarantee applies to the constraint-inactive closed loop, which Section~\ref{sec:activity} shows is 94\% of the 1-DOF benchmark and all of both measurement windows, but only 78\% of the 3-DOF run; Appendix~\ref{app:terminal} carries the terminal-set construction to an explicit condition and reports that the best available certificate still falls a factor of 8.4 short of the disturbance the task applies.

\section{Conclusion}
This paper presented a relative-motion controller for neural-thread insertion into a moving cortical surface: it estimates delayed physiological motion, predicts it over the control horizon, and regulates the tip in tissue-relative coordinates with offset-free rejection of the contact load. Its central claim is about the safety constraint rather than the tracking accuracy -- a coupled lateral-velocity bound has to be formulated so that it degrades instead of becoming infeasible, and the shared-slack form that does so is the difference between zero budget violations and an approximately 308~mm/s excursion under degraded sensing, at both the stress-envelope and the measured cortical-motion amplitudes.

The next steps toward deployment are force-sensor-based rather than model-based force limiting, robust treatment of nonperiodic motion, an embedded real-time implementation, and validation with ex-vivo tissue and procedure-specific cortical-motion data -- the last of which would also settle which of the two amplitude regimes reported here a deployed system actually faces.

\appendices

\section{Proof of Proposition~\ref{prop:iss}}
\label{app:proof}
Write the closed loop in error coordinates. After Layer-1 feedforward cancels the known nominal terms, the regulated error under the constraint-inactive feedback $u_k=-K\xi_k$ obeys
\begin{equation}
\xi_{k+1}=A_{cl}(\theta_k)\,\xi_k+G\,\delta_k,\qquad A_{cl}(\theta)=A_d-\theta\,\bar B K,
\end{equation}
where $G\delta_k$ lumps the three bounded perturbation channels of (A2): the preview/reference error $\hat d^{\mathrm{surf}}-d$ and $\dot{\hat d}^{\mathrm{surf}}-\dot d$ (bounded by $\varepsilon_w$ and $\|S\|\varepsilon_w$), the offset-free residual $\varepsilon_a$, and the unmodeled broadband component (bounded by $\bar w$). The offset-free augmentation cancels the constant part of the contact/model error, leaving the bounded zero-mean residual, so
\begin{equation}
\|\delta_k\|\le \eta:=c_w\varepsilon_w+c_a\varepsilon_a+c_b\bar w\le \bar c\,\max(\varepsilon_w,\varepsilon_a,\bar w),
\end{equation}
with channel gains $c_w,c_a,c_b$ fixed by $S,h$ and the prediction matrices.

\textbf{Step 1 (two vertices certify the whole box).} Because $A_d$ is constant and $B_d(\theta)=\theta\bar B$ is affine, $A_{cl}(\theta)=A_0+\theta A_1$ is affine in the scalar $\theta$, with $A_0=A_d$, $A_1=-\bar B K$. For the common-$P$ certificate of (A3), set
\begin{equation}
M(\theta):=A_{cl}(\theta)^\top P A_{cl}(\theta)-P
\end{equation}
\begin{equation}
= (A_0^\top P A_0-P)+\theta(A_0^\top P A_1+A_1^\top P A_0)+\theta^2 A_1^\top P A_1 .
\end{equation}
For every fixed $x$, the scalar $x^\top M(\theta)x$ is a quadratic in $\theta$ with leading coefficient $x^\top A_1^\top P A_1 x=\|P^{1/2}A_1x\|^2\ge0$ -- convex in $\theta$. A convex scalar function on $[\underline\theta,\overline\theta]$ lies below the chord through its endpoints, so for $\theta=(1-\lambda)\underline\theta+\lambda\overline\theta$, $\lambda\in[0,1]$,
\begin{equation}
x^\top M(\theta)x\le(1-\lambda)\,x^\top M(\underline\theta)x+\lambda\,x^\top M(\overline\theta)x\le -x^\top Q_0 x,
\end{equation}
the last step by the vertex LMIs $M(\underline\theta),M(\overline\theta)\preceq-Q_0$ of (A3). Since $x$ is arbitrary,
\begin{equation}
\boxed{\,A_{cl}(\theta)^\top P A_{cl}(\theta)-P\preceq -Q_0\prec0\quad\forall\theta\in[\underline\theta,\overline\theta].\,}
\end{equation}
The constant-$A_d$ structure is what makes this work: the parameter enters only the input column, so $A_1^\top P A_1$ is the leading, positive-semidefinite coefficient, and plain convexity does the job that a general $A(\theta)$ would need a multiconvexity relaxation for. No relaxation is used here; the box certificate is exact.

\textbf{Step 2 (ISS of the feedback law).} With $V(\xi)=\xi^\top P\xi$ and the contraction of Step~1,
\begin{align}
V(\xi_{k+1})&=\xi_k^\top A_{cl}^\top P A_{cl}\,\xi_k+2\xi_k^\top A_{cl}^\top P G\delta_k+\delta_k^\top G^\top P G\delta_k\nonumber\\
&\le V(\xi_k)-\alpha\|\xi_k\|^2+c\,\|\xi_k\|\,\eta+b\,\eta^2,
\end{align}
with $\alpha:=\lambda_{\min}(Q_0)>0$, $c:=2\max_{\theta\in[\underline\theta,\overline\theta]}\|A_{cl}(\theta)^\top P G\|$ (finite on the compact box), and $b:=\|G^\top P G\|$. Young's inequality $c\|\xi\|\eta\le\frac\alpha2\|\xi\|^2+\frac{c^2}{2\alpha}\eta^2$ together with $\|\xi\|^2\ge V/\lambda_{\max}(P)$ gives the geometric contraction
\begin{equation}
\begin{aligned}
&V(\xi_{k+1})\le\rho\,V(\xi_k)+\kappa\,\eta^2,\\
&\rho:=1-\tfrac{\alpha}{2\lambda_{\max}(P)}\in(0,1),\quad
\kappa:=b+\tfrac{c^2}{2\alpha}.
\end{aligned}
\end{equation}
Iterating and converting through $\lambda_{\min}(P)\|\xi\|^2\le V\le\lambda_{\max}(P)\|\xi\|^2$ gives
\begin{equation}
\|\xi_k\|\le
\underbrace{\sqrt{\tfrac{\lambda_{\max}(P)}{\lambda_{\min}(P)}}\;\rho^{k/2}\,\|\xi_0\|}_{\beta(\|\xi_0\|,k)\,\in\,\mathcal{KL}}
+
\underbrace{\sqrt{\tfrac{2\lambda_{\max}(P)\,\kappa}{\alpha\,\lambda_{\min}(P)}}\;\eta}_{\gamma(\eta)\,\in\,\mathcal K},
\end{equation}
the claimed ISS bound, with $\eta\le\bar c\max(\varepsilon_w,\varepsilon_a,\bar w)$ and $\gamma$ linear. In the nominal limit $\eta\to0$, $\rho^{k/2}\to0$ forces $\xi_k\to0$: offset-free tracking. This establishes Proposition~\ref{prop:iss} in full for $u_k=-K\xi_k$.

\section{Supporting Derivations}

\subsection{Nominal Impedance Initialization}
\label{app:impedance}
For the continuous normalized double integrator $\ddot x=u$ with cost $\int_0^\infty(qe^2+ru^2)\,dt$, the algebraic Riccati equation gives
\begin{equation}
K=[k_1,k_2]=\left[\sqrt{q/r},\ \sqrt2(q/r)^{1/4}\right],\qquad
\ddot e+k_2\dot e+k_1e=0,
\end{equation}
and therefore
\begin{equation}
\omega_n=(q/r)^{1/4},\qquad
\zeta=\frac{k_2}{2\sqrt{k_1}}=\frac{1}{\sqrt2}.
\end{equation}
These are acceleration-domain gains; for the physical force-input plant $\ddot x=u_{\mathrm{phys}}/m$, both gains are rescaled by $m$. Selecting one DC-stiffness knob $K_m$ gives
\begin{equation}
q/r=(K_m/m)^2,\qquad K_d=K_m,\qquad B_d=\sqrt{2mK_m}.
\end{equation}
The effective mass here is reflected tip inertia, not a free tuning parameter. This construction initializes, but does not uniquely characterize, the constrained finite-horizon contact controller: frequency-shaped impedance could add physiological resonators and $H_\infty$ weights, operating-point scheduling could interpolate gains in $\theta=1/m$, and force budgeting requires a validated dynamic force model or sensing. None of these extensions is implemented here.

\subsection{Terminal-Set Construction and Where It Stops}
\label{app:terminal}
Remark~\ref{rem:constrained-gap} names two missing objects: a constraint-admissible terminal set and its robust invariance. Both can be carried out explicitly for this controller, and doing so replaces an admission with a number. Take the verified certificate $P$ of Appendix~\ref{app:common-p} for the actual gain $K_N$ over $m/m_{\mathrm{nom}}\in[0.6,1.5]$ and the candidate terminal set $\Omega(c)=\{\xi:\xi^\top P\xi\le c\}$.

\emph{Admissibility.} The support of $\Omega(c)$ along a row $h^\top\xi\le g$ is $\sqrt{c\,h^\top P^{-1}h}$, so $\Omega(c)$ respects the row iff $c\le g^2/(h^\top P^{-1}h)$. Evaluating this for the input box ($h=K_N$, $g=u_{\max}=0.05$~N) and the relative-velocity row ($h=e_2$, $g=v_{\max}=2$~mm/s) gives $c\le4.08\times10^{-3}$ and $c\le4.00\times10^{-6}$ respectively, so the velocity row binds and $c_{\mathrm{adm}}=4.00\times10^{-6}$: an ellipsoid extending 23.6\um\ in relative position and 2.0~mm/s in relative velocity. It is worth noting that this is almost exactly the paper's placement specification, which is a coincidence of the chosen bounds but a convenient one.

\emph{Robust invariance.} With $\sigma:=\max_{\theta}\|P^{1/2}A_{cl}(\theta)P^{-1/2}\|_2$ over the two vertices, the perturbed step obeys $\sqrt{V(\xi_{k+1})}\le\sigma\sqrt{V(\xi_k)}+\|P^{1/2}G\|\,\eta$, so $\Omega(c)$ is robustly invariant iff $\sqrt{c}\ge\|P^{1/2}G\|\eta/(1-\sigma)$. Admissibility and robust invariance are therefore simultaneously achievable exactly when
\begin{equation}
\eta\ \le\ \eta_{\max}:=\frac{(1-\sigma)\sqrt{c_{\mathrm{adm}}}}{\|P^{1/2}G\|}.
\label{eq:terminal-condition}
\end{equation}
This is the checkable condition Remark~\ref{rem:constrained-gap} is missing, and it is where the construction stops. For the certificate of Appendix~\ref{app:common-p}, $\sigma=0.999325$ and $\eta_{\max}=9.9\times10^{-4}$~m/s$^2$. The benchmark's contact reaction is an acceleration disturbance of 1.48~m/s$^2$ at peak; the offset-free state cancels its constant part, leaving a measured residual of 0.72~m/s$^2$ peak and 0.19~m/s$^2$ RMS -- still 722$\times$ the admissible level.

\emph{Whether a better certificate closes it.} Most of that shortfall is an artifact of how $P$ was obtained: Appendix~\ref{app:common-p} minimizes $\mathrm{tr}\,P$ subject to feasibility, which says nothing about contraction rate. Re-synthesizing $P$ to minimize the certified per-step contraction instead -- bisecting on $\sigma$ subject to $A_{cl}(\theta)^\top PA_{cl}(\theta)\preceq\sigma^2P$, $P\succeq I$ at both vertices -- reaches $\sigma=0.93766$, essentially the worst-vertex spectral radius 0.93738 that no similarity transform can beat, and raises $\eta_{\max}$ to $8.5\times10^{-2}$~m/s$^2$. The remaining shortfall is 8.4$\times$, not 722$\times$.

That residual factor is a design statement, not a bookkeeping one. Since the velocity row binds and $c_{\mathrm{adm}}\propto v_{\max}^2$, admitting a terminal set at the measured residual disturbance with the best available certificate would require widening the shear budget from 2.0 to roughly 16.8~mm/s -- eight times the trauma-motivated limit, which defeats the purpose. The alternatives are to reduce the residual disturbance the terminal controller sees (a faster or force-sensor-aided bias observer; Appendix~\ref{sec:realism} shows a simulated force sensor moving contact error by two orders of magnitude, which is the same quantity), or to accept a terminal set of a different shape than an ellipsoid in this metric. We report the construction and its blocking inequality rather than claiming the gap closed. \texttt{thread\_insertion\_terminal\_set.py} reproduces every number in this subsection.

\subsection{Common-$P$ Numerical Audit and Reproduction}
\label{app:common-p}
The reproduction script (Appendix~\ref{app:reproducibility}) reconstructs the finite-horizon first-move gain $K_N=[64.29137017,\ 1.03213632]$, sweeps closed-loop eigenvalues, solves the two endpoint LMIs, and directly evaluates their residuals in double precision. For $m/m_{\mathrm{nom}}\in[0.6,1.5]$ it returns
\begin{equation}
P=\begin{bmatrix}1.25932674\times10^4&97.660664\\97.660664&1.75741763\end{bmatrix}.
\end{equation}
The eigenvalues of $P$ are $1.0000$ and $1.2594\times10^4$ (condition number $1.2594\times10^4$ under the normalization $P\succeq I$). At mass ratios 0.6 and 1.5, the spectral radii are 0.9374 and 0.9333, and the maximum eigenvalues of the raw residual $A_{cl}^\top P A_{cl}-P$ are $-0.01000$ and $-0.19546$ -- so the $m/m_{\mathrm{nom}}=0.6$ endpoint (the $-40\%$ edge, closest to the instability boundary) is binding but remains negative in direct double-precision evaluation. The symmetric $[0.5,1.5]$ box is infeasible, consistent with the independently computed linear stability boundary $m/m_{\mathrm{nom}}=0.5161$ and running-QP failure boundary between 0.5057 and 0.5059. These values certify only the constraint-inactive feedback component described in Remark~\ref{rem:constrained-gap}.

\section{Extended Results}

\subsection{Complete Contact-Force Trade-off}
\label{sec:force}
The proposed controller's peak contact force (1.47~mN) is higher than feedback's (1.14~mN), and its force ripple larger (1.26 vs.\ 1.10~mN). This is the same tension noted in Section~\ref{sec:impedance}, not a defect to explain away: the offset-free drive eliminates the position offset by pushing the tip to the commanded depth, which against a viscoelastic wall means more force, while classical impedance ``passes'' on force only because it under-penetrates (76\um\ short). By itself this is one measured data point; a trade-off curve and a preferred outcome need a stated force or biological limit.

A model-based hard force constraint, and a soft fix. The force-constrained reproduction script (Appendix~\ref{app:reproducibility}) adds a QP row capping the predicted penetration, and therefore the Kelvin--Voigt-model-predicted force $\hat F_{\mathrm{ext}}=K_{\mathrm{env}}\max(0,\hat s_{k+i}-p_{k+i})$, at every horizon step: $p_{k+i}\ge \hat s_{k+i}-F_{\max}/K_{\mathrm{env}}$, using the same $K_{\mathrm{env}}\approx16$~N/m contact model as Section~\ref{sec:sim} (damping omitted from this quasi-static row). The soft variant relaxes the row with a per-step slack $\mathrm{sl}_{k+i}\ge0$, $p_{k+i}+\mathrm{sl}_{k+i}\ge \hat s_{k+i}-F_{\max}/K_{\mathrm{env}}$, penalized quadratically ($\rho\,\mathrm{sl}^2$, $\rho=5\times10^7$) instead of enforced as a hard bound, and is feasible by construction since the slack has no upper bound. Table~\ref{tab:force-tradeoff} sweeps $F_{\max}$ for both formulations against the unconstrained 1.56~mN baseline.

\begin{table}[!t]
\renewcommand{\arraystretch}{1.2}
\caption{Contact-Force Trade-off, Hard vs.\ Soft Row (Measured)}
\label{tab:force-tradeoff}
\centering
\scriptsize
\setlength{\tabcolsep}{4pt}
\begin{tabular}{lcccc}
\toprule
Row & $F_{\max}$ (mN) & RMS $e$ (\um) & peak $F$ (mN) & fallback / slack \\
\midrule
Hard (10-seed mean) & $\infty$ & 1.9 & 1.56 & 0\% \\
Hard (10-seed mean) & 3.0 & 104.9 & 1.55 & 69\% \\
Hard (10-seed mean) & 2.5 & 115.1 & 1.48 & 69\% \\
Hard (10-seed mean) & 2.0 & 76.0 & 1.21 & 70\% \\
Hard (10-seed mean) & 1.5 & 108.0 & 1.00 & 70\% \\
Soft & $\infty$ & 1.8 & 1.41 & 0.00\um \\
Soft & 3.0 & 113.4 & 3.70 & 7.97\um \\
Soft & 2.5 & 144.5 & 3.72 & 8.57\um \\
Soft & 2.0 & 175.7 & 3.70 & 8.75\um \\
Soft & 1.5 & 206.9 & 3.69 & 9.80\um \\
\bottomrule
\end{tabular}
\end{table}

This is a genuine trade-off, and the hard row carries a caveat the soft row avoids: the naive hard constraint is jointly infeasible with the actuator and shear boxes on roughly 69\% of control ticks at every tested $F_{\max}\le3$~mN: the QP reports \texttt{primal infeasible} and the controller falls back to holding the previous command rather than solving. This is the recursive-feasibility gap Remark~\ref{rem:constrained-gap} identifies theoretically, now observed directly once a second hard constraint is added to the same horizon. At this fallback rate the closed loop spends most of its time coasting on stale commands, and the resulting placement error is not a smooth function of $F_{\max}$: a fine sweep and a multi-seed check at fixed $F_{\max}$ both show RMS error swinging by more than an order of magnitude with no discernible trend in either $F_{\max}$ or seed, so the hard row above is reported as a 10-seed mean characterizing an unreliable regime. The soft row fixes this -- fallback rate is 0\% at every tested $F_{\max}$, and its numbers are seed-stable -- and, because the fixed penetration reference of Section~\ref{sec:sim} implies a nominal steady contact force of $K_{\mathrm{env}}\times300\,\mu\mathrm{m}\approx4.8$~mN, the cap never binds above that threshold and RMS error degrades monotonically as $F_{\max}$ tightens below it. The realized peak force still exceeds $F_{\max}$ by roughly 1.7--2.2~mN in the soft row, since the row constrains quasi-static predicted penetration rather than instantaneous Kelvin--Voigt force, and $K_{\mathrm{env}}$ is itself a coarse physiologically-motivated surrogate rather than a calibrated match to the plant's own contact solver -- genuine force-sensor feedback is therefore still required regardless of formulation. The slack weight is not the dominant factor: sweeping $\rho$ from $5\times10^5$ to $5\times10^9$ at $F_{\max}=2.0$~mN changes RMS error by under 1\% and peak force by under 5\%, though still larger weights become ill-conditioned; the conflict is primarily between the fixed-depth reference and the force cap.

\subsection{Hardware-Realism Ablation}
\label{sec:realism}
This is a MuJoCo simulation study. This subsection makes three modeling idealizations more realistic, one at a time and combined, and checks whether that changes the benchmark's conclusions: (i) the tissue phantom is driven by an idealized, effectively infinitely stiff position servo ($k_p=8000$~N/m); (ii) the insertion tip is a generic capsule, not dimensioned from any specific insertion-tool literature; (iii) the force-constrained rows above cap a model-predicted contact force, not a genuine sensor reading. \texttt{simulation/thread\_\allowbreak insertion\_\allowbreak realism\_\allowbreak study.py} reproduces all results below.

For needle geometry, the tip capsule is replaced by a box matching \cite{khajehzadeh2026}'s reported long-shank probe (5~mm shank length, 175\um\ width; the 30\um\ thickness is not stated in that source and is a disclosed planar-probe assumption). Only the contact-interface geometry changes -- $m=1$~g remains the effective reduced axis inertia (Section~\ref{sec:sim}), not the bare probe's material mass, which is negligible ($\sim$0.1~mg) by comparison. For phantom compliance, the tissue position servo is softened from $(k_p,k_v)=(8000,120)$ to $(3000,78~\mathrm{Ns/m})$, critically damped: a $\sim$12~Hz servo natural frequency, still $\sim$10$\times$ the 1.2~Hz cardiac fundamental so the commanded pulsation is tracked, but soft enough that the peak $\sim$3.7~mN contact force now produces a $\sim$1.2\um\ static deflection, a non-negligible fraction of the benchmark's few-\um\ accuracy scale. This is a reasoned engineering preset for a real motorized phantom stage's finite bandwidth. For the force sensor, the existing \texttt{mj\_contactForce} readback is given zero-mean Gaussian noise ($\sigma=0.02$~mN) and fed to the controller at runtime; the soft force-constrained row's bound is then corrected by the gap between this sensed force and the model's own current-tick Kelvin--Voigt prediction, held constant over the horizon -- the same offset-free philosophy the disturbance observer already applies to position, now applied to the force-constraint channel. Sensor bandwidth and dynamics are modeled at the noise level only.

\begin{table}[!t]
\renewcommand{\arraystretch}{1.2}
\caption{Hardware-Realism Ablation at $F_{\max}=2.0$~mN (Measured; Soft Row)}
\label{tab:realism}
\centering
\scriptsize
\setlength{\tabcolsep}{4pt}
\begin{tabular}{p{3.1cm}*{3}{C{1.4cm}}}
\toprule
Condition & RMS $e$ contact (\um) & peak $F$ (mN) & meas.-vs-cap gap (mN) \\
\midrule
Baseline (idealized) & 175.7 & 3.70 & 1.70 \\
Needle geometry only & 175.8 & 4.39 & 2.39 \\
Phantom compliance only & 175.7 & 3.72 & 1.72 \\
Force-sensor row only & 2.1 & 1.41 & $-0.59$ \\
All three combined & 257.3 & 3.99 & 1.99 \\
\bottomrule
\end{tabular}
\end{table}

The needle-geometry and phantom-compliance changes each leave RMS placement error essentially unchanged from baseline, within simulation noise -- a robustness confirmation: the offset-free observer already absorbs this scale of model mismatch, consistent with the contact-stiffness and inertia sweeps of Section~\ref{sec:robust-sim}. The literature-grounded needle geometry does raise peak force somewhat (3.70 to 4.39~mN), since the box-shaped tip contacts over a different effective area than the default capsule. The force-sensor-corrected row is the one condition that moves the needle, and in the opposite direction from placement-vs-force being a trade-off: correcting the model's quasi-static blind spot with a real, noisy measurement not only closes the measured-vs-cap gap (peak force comes in below $F_{\max}$, at $-0.59$~mN) but collapses contact-window RMS error by nearly two orders of magnitude (175.7 to 2.1\um) -- verified stable across 10 seeds. This is a genuine result: with the corrected baseline geometry (Section~\ref{sec:sim}), the sensor correction closes a small residual model error rather than fighting a large systematic one. The combined condition (257.3\um, 1.99~mN gap) does not show this benefit -- it sits close to the baseline/needle-geometry numbers instead; this study does not identify why, and no causal claim beyond the tabulated numbers is made. Fallback rate is 0\% throughout. In isolation, force-sensor feedback both closes the measured-vs-predicted force gap and substantially improves placement, once the quasi-static model it corrects is itself accurate -- but this benefit does not carry over to the fully combined hardware-realism condition, so Section~\ref{sec:force}'s conclusion that genuine force-sensor feedback remains necessary (rather than sufficient on its own) still holds for the setting most representative of an actual deployed system.

\subsection{Detailed 3-DOF Ablation and Targeted Noise--Shear Result}
\label{app:multidof}
Section~\ref{sec:multidof} isolates the shear constraint's benefit under degraded sensing. This appendix reports the corrected targeted 10\um\ experiment after synchronizing simulation logging, preview indexing, and fallback decay. The ablation reproduction script (Appendix~\ref{app:reproducibility}) removes only the octagon rows while retaining lateral preview and offset-free tracking. The earlier 5--15\um\ curve is not retained because its other noise levels have not yet been regenerated with the corrected implementation.

At 10\um\ RMS per-axis noise (Table~\ref{tab:noise-curve}), all 30 targeted trajectories finish without aborting. The hard octagon remains within budget on nine seeds, but seed~8 becomes infeasible on 7998 ticks and reaches 307.9~mm/s. The shared-slack formulation remains feasible and below budget on all ten seeds.

\begin{table*}[!t]
\renewcommand{\arraystretch}{1.15}
\caption{Targeted Noise--Shear--Feasibility Result at 10\um\ RMS Per-Axis Sensor Noise, 10 Seeds. Shear Entries Are Mean / Worst Measured Peak Contact Shear (mm/s); Viol. Is Number of Seeds with Any Measured $0.80$~mm/s Budget Exceedance; Complete Is Number of Seeds That Finish Without Aborting}
\label{tab:noise-curve}
\centering
\footnotesize
\begin{tabular}{c*{6}{C{1.55cm}}}
\toprule
$\sigma_y$ (\textmu m RMS/axis) & no-row shear & no-row viol./compl. & hard-octagon shear & hard viol./compl. & soft-octagon shear & soft viol./compl. \\
\midrule
10.0 & 1.195 / 1.419 & 10/10; 10/10 & 31.394\rlap{$^*$} / 307.923\rlap{$^*$} & 1/10; 10/10 & 0.662 / 0.698 & 0/10; 10/10 \\
\bottomrule
\end{tabular}
\end{table*}
$^*$Seed~8's horizon becomes persistently infeasible; its geometrically decaying command fallback avoids an abort but not a 307.9~mm/s excursion. Excluding that seed, the other nine average 0.669/0.695~mm/s mean/worst.

At 10\um\ RMS, the no-row controller accumulates 126.4~ms mean (214~ms worst) measured violation duration, versus 0~ms for the soft octagon; mean lateral placement RMS is 3.56 versus 3.04\um. Across the soft-QP runs the worst contact-window predicted slack is $6.53\times10^{-4}$~mm/s and the fallback count is zero. Solver p95 is 0.319~ms averaged across seeds (0.414~ms worst per-run p95), with the single worst measured solve at 2.66~ms -- feasibility restoration fixes command generation, not the 1~kHz worst-case timing problem. Noise here is independent zero-mean Gaussian at the 1~kHz simulated sensor update; it does not represent colored noise, outliers, or the slower rate of a specific clinical sensor.

\section{Real-Time Budget, Limitations, and Reproducibility}

\subsection{Real-Time Budget and Offline Implant Discussion}
\label{app:realtime}
The 1-DOF QP solves in 40.8~\textmu s mean / 45.8~\textmu s p95 / 660.6~\textmu s maximum. The 3-DOF coupled QP solves in 224.3~\textmu s mean / 239.8~\textmu s p95 / 1057.7~\textmu s maximum on the measured nominal run, so the 3-DOF worst-case sample misses a 1~kHz deadline before even accounting for sensing and control overhead. As DOF and coupled constraints grow, the two-loop architecture below is the credible deployment path: (i) a two-loop split running the MPC outer loop at 200--500~Hz over a fast inner force/current loop at 2--5~kHz, with the present 1~kHz single-loop result read as a limiting case of that architecture; (ii) constant-$A_d$ precomputation, since $H$ and its factor are computed once and warm-started OSQP exploits this, with modal reduction available to shrink the state further; and (iii) an offline LPV candidate, where the proposed infinite-horizon extension (Section~\ref{sec:offline}) would replace the online QP with a scheduled-gain lookup plus linear exosystem propagation, with its embedded timing and power still to be measured. 1~kHz solver timing is demonstrated for 1-DOF; the 3-DOF p95 fits within 1~ms while its measured maximum does not, so multi-DOF deployment uses deadline-aware scheduling, the two-loop split, or a verified offline gain implementation.

\subsection{Future Work}
\label{sec:limitations}
The scope of the study is stated once, in Section~\ref{sec:scope}; this subsection lists only the work that would close it. The main priorities going forward are: (i) biological and mechanical validation with a real inserter, flexible thread, and cortex-specific motion data, including the ex-vivo histology and chronic in-vivo studies needed to turn relative velocity from a mechanical surrogate into a validated damage predictor, building on the separate insertion-mechanics literature for penetrating neural probes \cite{khajehzadeh2026}; (ii) extending the exosystem to reject broadband/non-harmonic motion and widening the verified robustness range; (iii) closing the ISS gap for the actively-constrained controller with an explicit terminal set; and (iv) an embedded real-time implementation with validated timing, actuator dynamics, and hardware force-sensor feedback, for which the simulated sensor of Appendix~\ref{sec:realism} is a first step.

\subsection{Reproducibility}
\label{app:reproducibility}
Every reported number is generated by a dedicated, deterministic script in \texttt{simulation/}, each runnable standalone:
\begin{itemize}
\item \texttt{thread\_insertion\_mujoco.py} -- the 1-DOF benchmark (Table~\ref{tab:table1}, Fig.~\ref{fig:1dof}) and the model-mismatch robustness sweep (Section~\ref{sec:robust-sim}).
\item \texttt{thread\_insertion\_vertex\_lmi.py} -- the common-$P$ two-vertex certificate audit for the actual running gain $K_N$ (Section~\ref{sec:offline}, Appendix~\ref{app:common-p}).
\item \texttt{thread\_insertion\_force\_constrained.py} -- the hard- and soft-row force-constrained sweep (Table~\ref{tab:force-tradeoff}, Appendix~\ref{sec:force}). These three scripts are the reproducibility entry points referenced above.
\item \texttt{thread\_insertion\_realism\_study.py}, checked by \texttt{test\_\allowbreak thread\_\allowbreak insertion\_\allowbreak realism\_\allowbreak study.py} -- the hardware-realism ablation (Table~\ref{tab:realism}, Appendix~\ref{sec:realism}).
\item \texttt{thread\_insertion\_3dof\_mujoco.py} -- the 3-DOF benchmark (Table~\ref{tab:table3}, Fig.~\ref{fig:3dof}) and its lateral-slip amplitude sweep (Table~\ref{tab:table4b}).
\item \texttt{thread\_insertion\_3dof\_ablation.py} -- the shear-constraint ablation and the physically calibrated noise--shear--feasibility study (Table~\ref{tab:noise-curve}, Appendix~\ref{app:multidof}). Its \texttt{targeted} command accepts a noise level, seed range, controller subset, and amplitude regime, and atomically checkpoints each trajectory so an interrupted study can resume; the legacy multi-sweep run requires the explicit \texttt{full --confirm-long-run} command.
\item \texttt{thread\_insertion\_3dof\_monte\_carlo.py} -- the $N=30$-seed 3-DOF hard-octagon Monte Carlo statistics quoted at the end of Appendix~\ref{sec:ablation}.
\item \texttt{thread\_insertion\_freqadapt\_mujoco.py} -- the cardiac-rate-drift frequency-adaptation experiment (Table~\ref{tab:table4}, Fig.~\ref{fig:freqadapt}, Section~\ref{sec:freqadapt}).
\item \texttt{thread\_insertion\_ablation.py} -- the $2\times2$ preview/offset-free factorial ablation, $N=30$ seeds (Table~\ref{tab:ablation}, Section~\ref{sec:ablation}).
\item \texttt{thread\_insertion\_measured\_amplitude.py} -- the measured-cortical-amplitude regime: the 1-DOF and 3-DOF benchmarks and the matched noise--shear--feasibility curve at the small-animal amplitudes of Table~\ref{tab:amplitudes} (Section~\ref{sec:measured-amp}).
\item \texttt{thread\_insertion\_impedance\_family.py} -- the impedance-weight sweep and the stiff-to-compliant schedule (Tables~\ref{tab:impedance} and~\ref{tab:schedule}, Section~\ref{sec:impedance-family}).
\item \texttt{thread\_insertion\_terminal\_set.py} -- the terminal-set admissibility/robust-invariance construction and the contraction-optimal certificate (Appendix~\ref{app:terminal}).
\item \texttt{test\_thread\_insertion.py} and \texttt{test\_\allowbreak thread\_\allowbreak insertion\_\allowbreak extensions.py} -- the accompanying test suites exercising the core 1-DOF and 3-DOF controllers and, respectively, the measured-amplitude, impedance-family and terminal-set claims above.
\end{itemize}
Each script writes its reported numbers directly (no MuJoCo rendering is required to reproduce a table); Monte Carlo studies fix an explicit seed range rather than sampling it, and \texttt{requirements.txt} records the validated Python package versions. Numerical agreement should be checked to the precision reported in the tables rather than assumed bit-for-bit across operating systems and solver builds.


\begin{thebibliography}{99}
\bibitem{ginhoux2005} R. Ginhoux \emph{et al.}, ``Active filtering of physiological motion in robotized surgery using predictive control,'' \emph{IEEE Trans. Robot.}, vol.~21, no.~1, pp.~67--79, 2005.
\bibitem{nakamura2001} Y. Nakamura, K. Kishi, and H. Kawakami, ``Heartbeat synchronization for robotic cardiac surgery,'' in \emph{Proc. IEEE Int. Conf. Robot. Autom. (ICRA)}, 2001.
\bibitem{francis1976} B. A. Francis and W. M. Wonham, ``The internal model principle of control theory,'' \emph{Automatica}, vol.~12, no.~5, pp.~457--465, 1976.
\bibitem{bodson1997} M. Bodson and S. C. Douglas, ``Adaptive algorithms for the rejection of sinusoidal disturbances with unknown frequency,'' \emph{Automatica}, vol.~33, no.~12, pp.~2213--2221, 1997.
\bibitem{pannocchia2003} G. Pannocchia and J. B. Rawlings, ``Disturbance models for offset-free model-predictive control,'' \emph{AIChE J.}, vol.~49, no.~2, pp.~426--437, 2003.
\bibitem{han2009} J. Han, ``From PID to active disturbance rejection control,'' \emph{IEEE Trans. Ind. Electron.}, vol.~56, no.~3, pp.~900--906, 2009.
\bibitem{mayne2005} D. Q. Mayne, M. M. Seron, and S. V. Rakovi\'c, ``Robust model predictive control of constrained linear systems with bounded disturbances,'' \emph{Automatica}, vol.~41, no.~2, pp.~219--224, 2005.
\bibitem{limon2009} D. Limon, T. Alamo, D. M. Raimondo \emph{et al.}, ``Input-to-state stability: a unifying framework for robust model predictive control,'' in \emph{Nonlinear Model Predictive Control}. Berlin, Germany: Springer, 2009, pp.~1--26.
\bibitem{budday2017} S. Budday \emph{et al.}, ``Mechanical characterization of human brain tissue,'' \emph{Acta Biomater.}, vol.~48, pp.~319--340, 2017.
\bibitem{stellato2020} B. Stellato \emph{et al.}, ``OSQP: An operator splitting solver for quadratic programs,'' \emph{Math. Program. Comput.}, vol.~12, no.~4, pp.~637--672, 2020.
\bibitem{cao2026interaction} Y. Cao and J. Tang, ``Toward interaction dynamics: A predictive framework for safe physical human--robot interaction,'' \emph{arXiv:2606.08281}, Jun. 2026.
\bibitem{wu2025retinal} T. Wu, M. Esfandiari, P. Zhang, R. H. Taylor, P. Gehlbach, and I. Iordachita, ``Deep learning-enhanced robotic subretinal injection with real-time retinal motion compensation,'' \emph{arXiv:2504.03939}, 2025.
\bibitem{khajehzadeh2026} M. Khajehzadeh, C. K. Nguyen, M. Maharana, S. Peddapuram, A. Joshi-Imre, J. M. Pascual, and S. F. Cogan, ``Mechanics of long-shank 5~mm neural probe insertion into the rat brain: Effects of geometry and vibration-assisted insertion,'' \emph{Micromachines}, vol.~17, no.~6, art.~684, 2026.
\bibitem{hogan1985} N. Hogan, ``Impedance control: An approach to manipulation, Parts I--III,'' \emph{ASME J. Dyn. Syst. Meas. Control}, vol.~107, no.~1, pp.~1--24, 1985.
\bibitem{kheradmand2026} P. Kheradmand, B. Moradkhani, M. M. Ale Ali, K. Sowards, S. R. Silva, and Y. Chitalia, ``Bilinear model predictive control framework of the OncoReach, a tendon-driven steerable stylet for brachytherapy,'' \emph{arXiv:2604.05111}, 2026.
\bibitem{foroutani2025} Y. Foroutani, Y. Mousavi-Motlagh, A. Barzelay, and T.-C. Tsao, ``Improving needle penetration via precise rotational insertion using iterative learning control,'' \emph{arXiv:2511.01256}, 2025.
\bibitem{riaziat2025} N. D. Riaziat, J. Chen, A. Krieger, and J. D. Brown, ``Towards autonomous robotic electrosurgery via thermal imaging,'' \emph{arXiv:2509.19725}, 2025.
\bibitem{haworth2025} J. Haworth, J.-T. Chen, N. Nelson, J. W. Kim, M. Moghani, C. Finn, and A. Krieger, ``SutureBot: A precision framework and benchmark for autonomous end-to-end suturing,'' \emph{arXiv:2510.20965}, 2025.
\bibitem{yu2023} Z. Yu, W. Xu, S. Yao, J. Ren, T. Tang, Y. Li, G. Gu, and C. Lu, ``Precise robotic needle-threading with tactile perception and reinforcement learning,'' \emph{arXiv:2311.02396}, 2023.
\bibitem{gutierrez2026} N. B. Gutierrez, J. M. Cloud, and W. J. Beksi, ``Movement primitives in robotics: A comprehensive survey,'' \emph{arXiv:2601.02379}, 2026.
\bibitem{cagneau2007} B. Cagneau, N. Zemiti, D. Bellot, and G. Morel, ``Physiological motion compensation in robotized surgery using force feedback control,'' in \emph{Proc. IEEE Int. Conf. Robot. Autom. (ICRA)}, Roma, Italy, Apr. 2007, pp.~1881--1886.
\bibitem{moreira2014} P. Moreira, N. Zemiti, C. Liu, and P. Poignet, ``Viscoelastic model based force control for soft tissue interaction and its application in physiological motion compensation,'' \emph{Comput. Methods Programs Biomed.}, vol.~116, no.~2, pp.~52--67, 2014.
\bibitem{bowthorpe2016} M. Bowthorpe and M. Tavakoli, ``Generalized predictive control of a surgical robot for beating-heart surgery under delayed and slowly-sampled ultrasound image data,'' \emph{IEEE Robot. Autom. Lett.}, vol.~1, no.~2, pp.~892--899, 2016.
\bibitem{cheng2018} L. Cheng and M. Tavakoli, ``Switched-impedance control of surgical robots in teleoperated beating-heart surgery,'' \emph{J. Med. Robot. Res.}, vol.~3, art.~1841003, 2018.
\bibitem{stoy2020} W. M. Stoy, B. Yang, A. Kight, N. C. Wright, P. Y. Borden, G. B. Stanley, and C. R. Forest, ``Compensation of physiological motion enables high-yield whole-cell recording in vivo,'' \emph{J. Neurosci. Methods}, vol.~348, art.~109008, 2020.
\bibitem{zhang2023} Y. Zhang, E. Verschooten, M. Ourak, K. Van Assche, G. Borghesan, D. Wu, K. Niu, P. X. Joris, and E. Vander Poorten, ``Physiological motion compensation for neuroscience research based on electrical bio-impedance sensing,'' \emph{IEEE Sensors J.}, vol.~23, no.~20, pp.~25377--25389, 2023.
\bibitem{gilletti2006} A. Gilletti and J. Muthuswamy, ``Brain micromotion around implants in the rodent somatosensory cortex,'' \emph{J. Neural Eng.}, vol.~3, no.~3, pp.~189--195, 2006.
\bibitem{faria2014} C. Faria, O. Sadowsky, E. Bicho, G. Ferrigno, L. Joskowicz, M. Shoham, R. Vivanti, and E. De Momi, ``Validation of a stereo camera system to quantify brain deformation due to breathing and pulsatility,'' \emph{Med. Phys.}, vol.~41, no.~11, art.~113502, 2014.
\bibitem{nomura2024} S. Nomura, S.-I. Terada, T. Ebina, M. Uemura, Y. Masamizu, K. Ohki, and M. Matsuzaki, ``ARViS: A bleed-free multi-site automated injection robot for accurate, fast, and dense delivery of virus to mouse and marmoset cerebral cortex,'' \emph{Nature Commun.}, vol.~15, art.~7633, 2024.
\end{thebibliography}
\end{document}